\documentclass[11pt,a4paper]{article}
\pdfoutput=1   
\usepackage[utf8]{inputenc}
\usepackage{tabularx}
\usepackage{alpha} 
\hypersetup{%
   pdftitle    = {Non-perturbative renormalisation and improvement of non-singlet tensor currents},
   pdfauthor   = {L.Chimirri, P.Fritzsch, J.Heitger, F.Joswig, M.Panero, C.Pena, D.Preti},
   pdfkeywords = {Lattice QCD, Nonperturbative effects, Renormalisation, O(a) improvement}
}%
\usepackage{lmodern}
\usepackage[outercaption]{sidecap}
\usepackage{xspace}
\usepackage{defs}
\usepackage{cleveref}
\usepackage{siunitx}
\begin{document}

\preprintno{\footnotesize%
HU-EP-23/51\\
IFT-UAM/CSIC-23-102\\
MS-TP-23-42\\
YITP-23-109
}

\title{%
Non-perturbative renormalisation and improvement \\
of non-singlet tensor currents in $\Nf=3$ QCD
}

\collaboration{\includegraphics[width=2.8cm]{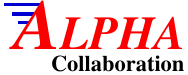}}

\author[hu,desy]{Leonardo~Chimirri}
\author[dub]{Patrick~Fritzsch}
\author[ms,yuk]{Jochen~Heitger}
\author[ed]{Fabian~Joswig}
\author[torinoPHYS,torinoINFN]{Marco~Panero}
\author[uam]{Carlos~Pena}
\author[torinoINFN]{David~Preti}

\address[hu]{Humboldt Universit\"at zu Berlin, IRIS Adlershof, Zum Gro{\ss}en Windkanal 6, 12489 Berlin, Germany}
\address[desy]{John von Neumann Institute for Computing (NIC), DESY, Platanenallee 6, 15738 Zeuthen, Germany}
\address[dub]{School of Mathematics, Trinity College Dublin, Dublin 2, Ireland}
\address[ms]{Institut f\"ur Theoretische Physik, Westf\"alische Wilhelms-Universit\"at M\"unster,\\ Wilhelm-Klemm-Stra{\ss}e 9, 48149 M\"unster, Germany}
\address[yuk]{Yukawa Institute for Theoretical Physics, Kyoto University, \\Kitashirakawa Oiwake-cho, Sakyo-Ku, Kyoto 606-8502, Japan}
\address[ed]{Higgs Centre for Theoretical Physics, School of Physics and Astronomy,\\ The University of Edinburgh, Edinburgh EH9 3FD, UK}
\address[torinoPHYS]{Department of Physics, University of Turin, Via Pietro Giuria 1, 10125 Turin, Italy}
\address[torinoINFN]{INFN, Sezione di Torino, Via Pietro Giuria 1, 10125 Turin, Italy}
\address[uam]{Departamento de F\'{\i}sica Te\'orica and Instituto de F\'{\i}sica Te\'orica UAM-CSIC,\\ Universidad Aut\'onoma de Madrid, Cantoblanco, 28049 Madrid, Spain}

\begin{abstract}
Hadronic matrix elements involving tensor currents play an important r\^ole in
decays that allow to probe the consistency of the Standard Model via precision
lattice QCD calculations. The non-singlet tensor current is a scale-dependent
(anomalous) quantity. We fully resolve its renormalisation group (RG) running
in the continuum by carrying out a recursive finite-size scaling technique.  In
this way ambiguities due to a perturbative RG running and matching to lattice
data at low energies are eliminated. We provide the total renormalisation
factor at a hadronic scale of 233 MeV, which converts the bare current into its
RG-invariant form.

Our calculation features three flavours of $\rmO(a)$ improved Wilson fermions
and tree-level Symanzik-improved gauge action. We employ the (massless)
Schr\"odinger functional renormalisation scheme throughout and present the
first non-perturbative determination of the Symanzik counterterm $\cT$ derived
from an axial Ward identity. We elaborate on various details of our
calculations, including two different renormalisation conditions.   
\\
\end{abstract}


\maketitle

\tableofcontents

\section{Introduction}\label{sec:basics}

The study of quantum chromodynamics (QCD), the fundamental theory of the strong
interaction, remains an active and very important area of research in
elementary particle theory. This is not only motivated by its most dramatic
phenomenological consequences (for example, the strong interaction directly
determines the largest part of the mass of baryons, and, as a consequence,
accounts for most of the mass of visible matter in the universe), but also by
the non-trivial r\^ole it plays in problems related to the physics of flavour,
including rare decays of heavy mesons (see, for instance,
Refs.~\cite{Buchalla:2008jp, Antonelli:2009ws, Bharucha:2015bzk, Blake:2016olu,
Cerri:2018ypt, Belle-II:2018jsg}), $\beta$-decays and the neutron electric
dipole moment~\cite{Bhattacharya:2015wna, Bhattacharya:2016zcn, Gupta:2018lvp,
Alexandrou:2019brg, FlavourLatticeAveragingGroupFLAG:2021npn}, and so on. Even
though these phenomena are determined by the electroweak interaction, the fact
that in the Standard Model (SM) the quarks carry both colour and electroweak
charges, and are confined within hadrons by the strong interaction, makes a
precise quantitative determination of the theoretical predictions of QCD a
crucial ingredient to test the SM against experimental results, with the
potential to disclose new-physics effects~\cite{Davoudi:2020ngi}. In view of
the negative results of direct searches for physics beyond the SM at the Large
Hadron Collider~\cite{LHCReinterpretationForum:2020xtr}, the motivation for
such tests is currently strongest than ever, since they could reveal the
existence of particles with masses beyond the reach of present collider
experiments.

The conventional framework to determine physical amplitudes involving hadronic
states is based on an effective weak Hamiltonian, whereby the effects of QCD
are encoded in matrix elements of effective quark field interactions. One type
of such interaction terms, which will be the main focus of the present work, is
given by flavour non-singlet bilinear quark currents with a tensor structure
for the Dirac indices:
\begin{align}
\label{tensor_current_continuum_definition}
T^a_{\mu\nu}(x) &= \mathrm{i}\bar{\psi}(x) \sigma_{\mu\nu} T^a \psi(x) \,,
\end{align}
where $\sigma_{\mu\nu}=\tfrac{\mathrm{i}}{2}[\gamma_{\mu},\,\gamma_{\nu}]$ acts
on the spinor indices, while $T^a$ is a generator of the $\mathrm{SU}(\Nf)$
group acting on the flavour indices.

It is important to note that, while partial-current-conservation laws protect
flavour non-singlet vector and axial currents from ultraviolet renormalisation,
tensor-like currents of the form~\eqref{tensor_current_continuum_definition}
are not constrained by such laws, and require an independent scale-dependent
renormalisation; in fact, the tensor current defined in
Eq.~\eqref{tensor_current_continuum_definition} is the only type of bilinear
operator whose evolution under renormalisation-group (RG) transformations
cannot be directly derived from that of quark masses. The anomalous dimension
associated with this current has been studied perturbatively both in continuum
schemes~\cite{Gracey:2000am, Almeida:2010ns}, where the most recent results
have been pushed to the four-loop order~\cite{Gracey:2022vqr}, and in lattice
schemes~\cite{Skouroupathis:2008mf}.

While perturbative expansions are reliable at high energies, the
non-perturbative character of the strong interaction at the scales typical of
hadrons requires an approach that does not rely on any weak-coupling
assumptions; this restricts the toolbox to derive the predictions of QCD for
processes taking place within hadronic states to numerical calculations in the
lattice regularisation (for an overview of the contributions lattice QCD can
give in the study of processes involving weak decays of heavy quarks and in the
refinement of SM predictions, see Refs.~\cite{Boyle:2022ncb,Boyle:2022uba}),
which is the formalism that we use here to study the renormalisation of the
tensor current. Among the different lattice discretisations for the Dirac
operator, the Wilson one~\cite{Wilson:1975id} turns out to be a particularly
convenient choice, since it offers various conceptual as well as practical
advantages. In particular, in the continuum limit (i.e., when the lattice
spacing $a$ is sent to zero) it completely removes the effects of all fermion
doublers, it is strictly ultralocal, and it explicitly preserves flavour
symmetry and the discrete symmetries of continuum QCD, while being
computationally much less demanding than other types of regularisations for the
Dirac operator (such as overlap or domain-wall fermions). The explicit
chiral-symmetry breaking introduced by the Wilson term, however, leads to
additive mass renormalisation for the quark fields and to the introduction of
discretisation effects at $\mathrm{O}(a)$, which reduce the convergence rate of
simulation results towards the continuum limit. This is an issue that affects
both the fermion action and the fermion
currents~\cite{Bochicchio:1985xa,Luscher:1996jn}, including the ones that are
the focus of this work. As will be discussed in detail in the present article,
this problem can be tackled in a systematic way by means of the Symanzik
improvement programme~\cite{Symanzik:1983dc,Symanzik:1983gh} and defining the
theory in the Schr\"odinger functional (SF) scheme~\cite{Luscher:1992an,
Luscher:1993gh, Sint:1993un} according to the framework presented in
Refs.~\cite{Sint:1997jx,Pena:2017hct}; this allows one to cancel the leading,
$\mathrm{O}(a)$, discretisation artifacts and to achieve $\mathrm{O}(a^2)$
scaling towards the continuum limit~\cite{Luscher:1996sc}. In particular, the
improvement of the fermion currents can be obtained by including additive
dimension-$4$ counterterms (which take the form of discretised derivatives of
vector currents), with appropriately tuned coefficients, in their definition on
the lattice.

This strategy, that here we apply on an ensemble of lattice configurations with
$\Nf=3$ dynamical quark flavours generated by the ALPHA
collaboration~\cite{DallaBrida:2016kgh,Campos:2018ahf}, is then expected to
yield the same level of non-perturbative control of the tensor current
renormalisation that has been previously obtained for the quark
masses~\cite{Capitani:1998mq, DellaMorte:2005kg, Campos:2018ahf} and to
contribute to the programme of non-perturbative improvement and renormalisation
for flavour-non-singlet quark field
bilinears~\cite{Capitani:1998mq,DellaMorte:2005kg,Blossier:2010jk,Bernardoni:2014fva,Bulava:2015bxa,
Bulava:2016ktf, DallaBrida:2018tpn, Fritzsch:2018zym, Gerardin:2018kpy,
Korcyl:2016ugy,Campos:2018ahf,Heitger:2021bmg,Heitger:2020mkp,Heitger:2020zaq}
and four-quark
operators~\cite{Guagnelli:2005zc,Palombi:2005zd,Dimopoulos:2006ma,Dimopoulos:2007ht,Palombi:2007dr,Papinutto:2014xna,Papinutto:2016xpq,Dimopoulos:2018zef}
pursued by the collaboration. For the error analysis, in the present work we
use the $\Gamma$-method approach~\cite{Wolff:2003sm, Schaefer:2010hu,
Ramos:2018vgu} as implemented in the \texttt{pyerrors} Python
package~\cite{Joswig:2022qfe}.

Non-perturbative renormalisation of tensor currents has been carried out in
$\text{RI}^\prime$-MOM schemes for many years, in simulations with different
numbers of dynamical quark flavours and using various types of
discretisations~\cite{Gockeler:1998ye,Becirevic:2004ny,Follana:2006rc,Aoki:2007xm,Sturm:2009kb,Constantinou:2010gr,Alexandrou:2012mt,Constantinou:2014fka,Hatton:2020vzp}.
The recent study in Ref.~\cite{Harris:2019bih}, also using RI-MOM, shares the
same lattice regularisation as the present work. To our knowledge, however,
this is the first instance of a non-perturbative computation of the
renormalisation group running of non-singlet tensor currents in the whole range
of energies relevant to SM physics.

The structure of this article is as follows. After discussing the pattern of
renormalisation and $\mathrm{O}(a)$ improvement of tensor currents in
Section~\ref{sec:RandI}, we present our non-perturbative calculation of the
tensor currents' improvement coefficient in Section~\ref{sec:improvement} and
the renormalisation of the tensor current in Section~\ref{sec:renormalisation}.
Our main findings are then summarised and discussed in
Section~\ref{sec:summary}. Finally, the Appendices include the covariance
matrices of our fits (Appendix~\ref{app:cov_matrices}), the detailed results
for the tensor-current improvement coefficient
(Appendix~\ref{app:improvement}), and a set of tables with the results for the
step scaling of the renormalisation factor (Appendix~\ref{app:running}).

Preliminary results of this work were presented in
Ref.~\cite{Chimirri:2019xsv}, while a more extensive discussion of the
framework of our calculation can be found in Ref.~\cite{Joswig:2021rcn}.

\section{Renormalisation and $\mathrm{O}(a)$ improvement of tensor currents}\label{sec:RandI}

Let $\mu$ denote the scale at which theory parameters and operators are
renormalised. The scale dependence of these quantities is given by their RG
evolution. The Callan--Symanzik equations satisfied by the gauge coupling and
quark masses are of the form  
\begin{align} \label{coupling_RGE}
  \mu\frac{\partial\gbar  }{\partial\mu} &= \beta(\gbar(\mu))            \,,\\\label{mass_RGE}
  \mu\frac{\partial\mbar_i}{\partial\mu} &= \tau(\gbar(\mu))\mbar_i(\mu) \,,
\end{align}
respectively, with renormalised coupling $\gbar$ and masses $\mbar_i$; the
index $i$ runs over flavour.  Starting from the renormalisation-group equation
(RGE) for correlation functions, we can also write the RGE for the insertion of
a multiplicatively renormalisable local composite operator $\mathcal{O}$ in an
on-shell correlator as
\begin{align}\label{oper_RGE}
        \mu\frac{\partial\overline{\mathcal{O}}(\mu)}{\partial\mu} &= \gamma_{\mathcal{O}}(\gbar(\mu))\overline{\mathcal{O}}(\mu) \,,
\end{align}
where $\overline{\mathcal{O}}(\mu)$ is the renormalised operator. The latter is
connected to the bare operator insertion $\mathcal{O}(g_0^2)$ through
\begin{align}\label{renormalizedO}
\overline{\mathcal{O}}(\mu) &= \lim_{a \to 0} Z_\mathcal{O}(g_0^2,a\mu) \mathcal{O}(g_0^2) \,,
\end{align}
where $g_0$ is the bare coupling, $Z_\mathcal{O}$ is a renormalisation factor,
and $a$ is some inverse ultraviolet cutoff --the lattice spacing in this work.
We assume a mass-independent scheme, such that both the $\beta$-function and
the anomalous dimensions $\tau$ and $\gamma_{\mathcal{O}}$ depend only on the
coupling and on the number of flavours $\Nf$ (other than on the number of
colours $\Nc$); examples of such schemes are the $\MSbar$ scheme of dimensional
regularisation~\cite{tHooft:1973mfk,Bardeen:1978yd}, RI
schemes~\cite{Martinelli:1994ty}, or the SF schemes we shall use to determine
the running non-perturbatively~\cite{Luscher:1992an,Jansen:1995ck}.  The RG
functions then admit asymptotic expansions of the form:
\begin{align}\label{eq:asympt-beta}
  \beta(g) \ &\overset{g \rightarrow 0}{\sim} \
  -g^3\big(b_0 + b_1 g^2 + b_2 g^4 + \ldots \big) \,, \\\label{eq:asympt-tau}
  \tau(g) \ &\overset{g \rightarrow 0}{\sim} \
  -g^2\big(d_0 + d_1 g^2 + d_2 g^4 + \ldots \big) \,, \\\label{eq:asympt-gamma}
  \gamma_{\mathcal{O}}(g) \ &\overset{g \rightarrow 0}{\sim} \
  -g^2\big(\gamma_{\mathcal{O}}^{(0)} + \gamma_{\mathcal{O}}^{(1)} g^2 + \gamma_{\mathcal{O}}^{(2)} g^4 + \ldots \big) \,.
\end{align}
The coefficients $b_0$, $b_1$ and $d_0$, $\gamma_{\mathcal{O}}^{(0)}$ are
independent of the renormalisation scheme chosen. In
particular~\cite{Vanyashin:1965ple,Khriplovich:1969aa,tHooft:1971akt,Gross:1973id,Politzer:1973fx,Caswell:1974gg,Jones:1974mm},
we have
\begin{align}
  b_0 &= \frac{1}{(4\pi)^2}\left ( \frac{11}{3}\Nc-\frac{2}{3}\Nf \right ) \,, \\
  b_1 &= \frac{1}{(4\pi)^4}\left [ \frac{34}{3}\Nc^2-\left ( \frac{13}{3}\Nc-\frac{1}{\Nc} \right ) \Nf \right ]  \,,
\end{align}
and
\begin{gather}
  d_0 = \frac{6\CF}{(4\pi)^2}\,,
\end{gather}
where $\CF=\frac{\Nc^2-1}{2\Nc}$ is the eigenvalue of the quadratic Casimir
operator for the fundamental representation of the algebra of the
$\mathrm{SU}(\Nc)$ gauge group, i.e., $\CF=\frac{4}{3}$ in QCD with three
colours.

The RGEs~(\ref{coupling_RGE}--\ref{oper_RGE}) can be formally solved in terms
of the renormalisation-group invariants (RGIs) $\Lambda_{\rm\scriptscriptstyle
QCD}$, $\hat{m}_i$ and $\hat{\mathcal{O}}$, respectively, as:%
\footnote{Our choice for the normalisation of $\hat{m}_i$ follows Gasser and
          Leutwyler~\cite{Gasser:1982ap,Gasser:1983yg,Gasser:1984gg}, whereas
          for Eq.~\eqref{rgi_operator} we have chosen the most usual
          normalisation with a power of $\alpha_{\rm s}$.
}
\begin{align}
  \label{rgi_coupling}
  \Lambda_{\rm\scriptscriptstyle QCD} &\,=\, \mu \,
  \frac{[b_0\gbar^2(\mu)]^{-b_1/2b_0^2}}{e^{1/2b_0\gbar^2(\mu)}} \,
    \exp\left\{-\int_0^{\gbar(\mu)}\dif g
      \left[\frac{1}{\beta(g)}+\frac{1}{b_0g^3}-\frac{b_1}{b_0^2g}\right]\right\}
    \,, \\
  \label{rgi_mass}
  \hat{m}_i &\,=\, \mbar_i(\mu) \, [2b_0\gbar^2(\mu)]^{-d_0/2b_0} \,
    \exp\left\{-\int_0^{\gbar(\mu)}\dif g
      \left[\frac{\tau(g)}{\beta(g)}-\frac{d_0}{b_0g}\right]\right\}
    \,, \\
  \nonumber
  \hat{\mathcal{O}} &\,=\, \overline{\mathcal{O}}(\mu) \, \left[\frac{\gbar^2(\mu)}{4\pi}\right]^{-\gamma_{\mathcal{O}}^{(0)}/2b_0} \,
    \exp\left\{-\int_0^{\gbar(\mu)}\dif g \,
      \left[\frac{\gamma_{\mathcal{O}}(g)}{\beta(g)}-\frac{\gamma_{\mathcal{O}}^{(0)}}{b_0g}\right]\right\}\\
  \label{rgi_operator}
    &\,\equiv\,\hat{c}(\mu)\overline{\mathcal{O}}(\mu)\,.
\end{align}
While the value of the $\Lambda_{\rm\scriptscriptstyle QCD}$ parameter depends
on the renormalisation scheme chosen, $\hat{m}_i$ and $\hat{\mathcal{O}}$ are
the same for all schemes. In this sense, they can be regarded as meaningful
physical quantities, as opposed to their scale-dependent counterparts. The aim
of the non-perturbative determination of the RG running of parameters and
operators is to connect the RGIs---or, equivalently, the quantity renormalised
at a very high energy scale, where perturbation theory can be applied---to the
bare parameters or operator insertions, computed in the hadronic energy regime.
In this way the three-orders-of-magnitude leap between the hadronic and weak
scales can be bridged without significant uncertainties related to the use of
perturbation theory.

In this work, we shall focus on the renormalisation of the tensor currents
introduced in Eq.~\eqref{tensor_current_continuum_definition}.
The universal one-loop coefficient of the tensor anomalous dimension is
\begin{gather}
  \gamT^{(0)}=\frac{2\CF}{(4\pi)^2} \,.
\end{gather}
In the two SF schemes we shall consider below, labelled by the superscripts \ff
and \kk, the two-loop anomalous dimension reads \cite{Pena:2017hct}
\begin{align}
        \gamT^{(1),\ff}&=0.0069469(8) - 0.00022415(5)\times \Nf\,,\\
        \gamT^{(1),\kk}&=0.0063609(8) - 0.00018863(5)\times \Nf\,,
\end{align}
where the numbers in parentheses represent the uncertainty on the last
significant figure.

As already done in the introduction, it is important to observe that the tensor
current is the only bilinear operator that evolves under RG transformation in a
different way than quark masses---whereas partial conservation of the vector
and axial currents protects them from renormalisation, and fixes the anomalous
dimension of both scalar and pseudoscalar densities to be $-\tau$.

So far we have discussed the formal renormalised continuum theory. In practice,
renormalisation is worked out by first introducing a suitable regulator, which
in our case will be a spacetime lattice with Wilson fermion action for quark
fields. This implies leading cutoff effects of $\mathrm{O}(a)$, which can be
reduced down to $\mathrm{O}(a^2)$ by implementing Symanzik's improvement
programme.  This requires both adding the Sheikholeslami--Wohlert
term~\cite{Sheikholeslami:1985ij} to the fermion action, and appropriate
dimension-$4$ counterterms to fermion currents, with coefficients tuned so as
to cancel $\mathrm{O}(a)$ contributions.  In the case of the flavour
non-singlet tensor currents~\eqref{tensor_current_continuum_definition}, the
only improvement term surviving the chiral limit has the form
\begin{align}
\label{eq:improved_tensor_current}
  \big(T_{\mu\nu}^a\big)^\mathrm{I}(x) &=
       T_{\mu\nu}^a(x)+a\cT\big(\tilde{\partial}_\mu V_\nu^a(x)-\tilde{\partial}_\nu V_\mu^a(x)\big) \,,
\end{align}
where the flavour non-singlet local vector current is defined as
\begin{align}
   V^a_{\mu}(x) &= \bar{\psi}(x)\gamma_\mu T^a \psi(x) \,.
\end{align}
The improvement coefficient $\cT$ was determined at one-loop order in
perturbation theory for the Wilson gauge action
in Refs.~\cite{Sint:1997jx,Pena:2017hct}
\begin{align}
	\cT^\mathrm{1-lp}(g_0^2) &= 0.00896(1)\times \CF\, g_0^2 \,,
\end{align}
while for the L\"uscher--Weisz gauge action one has~\cite{Taniguchi:1998pf}
\begin{align}
	\cT^\mathrm{1-lp}(g_0^2) &= 0.00741 \times \CF \, g_0^2 \,.
\end{align}

As in the remainder of this work all calculations are performed at zero
momentum and we always sum over spatial components,
only the chromoelectric components require the improvement term, while the
chromomagnetic ones are automatically $\mathrm{O}(a)$ improved, viz.
\begin{align}
 \sum_{\mathbf{x}}\big(T_{0k}^a\big)^\mathrm{I}(x)&=\sum_{\mathbf{x}}\big(T_{0k}^a(x)+a\cT\tilde{\partial}_0 V_k^a(x)\big)\,,\\
 \sum_{\mathbf{x}}\big(T_{ij}^a\big)^\mathrm{I}(x)&=\sum_{\mathbf{x}}T_{ij}^a(x)\,. \label{eq:chromo_magnetic_improvement}
\end{align}
Renormalised tensor currents in the continuum limit can then be obtained from
bare $\mathrm{O}(a)$ improved currents as, e.g.,
\begin{gather}\label{eq:renormalized_current}
\overline{T}^a_{0k}(\mu) = \lim_{a\to 0}\ZT(g_0^2,a\mu)(T^a_{0k})^\mathrm{I}(g_0^2)\,,
\end{gather}
where $\ZT$ is the renormalisation factor obtained from some suitable
renormalisation condition and $T^a_{0k}(g_0^2)$ is a shorthand notation for the
insertion of the tensor current in a bare correlation function computed at bare
coupling $g_0^2$.

In the next two Sections, we shall discuss the non-perturbative determination
of the $\mathrm{O}(a)$ improvement coefficient $\cT$ and the renormalisation
constant $\ZT$, to carry out the computation of non-perturbatively renormalised
tensor currents in the whole range of scales of interest for SM physics.  For
the computation of $\cT$ and $\ZT$ we shall employ a SF
setup~\cite{Luscher:1992an,Sint:1993un}, for which we shall adopt the
conventions and notations introduced in Ref.~\cite{Luscher:1996sc}.  The SF
framework amounts to formulating QCD in a finite space-time volume of size $L^3
\times T$, with inhomogeneous Dirichlet boundary conditions at Euclidean times
$x_0=0$ and $x_0=T$. The boundary condition for gauge fields has the form
\begin{gather}
\left.U_k(x)\right|_{x_0=0}=\mathcal{P}\exp\left\{
a\int_0^1\dif t \,\,C_k({\mathbf{x}}+(1-t)a\hat{\mathbf{k}})
\right\}\,,
\end{gather}
where $\hat{\mathbf{k}}$ is a unit vector in the direction $k$,
$\mathcal{P}\exp$ denotes a path-ordered exponential, and $C_k$ is some smooth
gauge field. A similar expression applies at $x_0=T$ in terms of another field
$C'_k$.  Fermion fields obey the boundary conditions
\begin{align}
      P_+\psi(x)\Big|_{x_0=0}\hspace*{-0.8em} &= \rho({\mathbf{x}})             \,,&
  \bar\psi(x)P_-\Big|_{x_0=0}\hspace*{-0.8em} &= \bar\rho({\mathbf{x}})         \,,&
      P_-\psi(x)\Big|_{x_0=0}\hspace*{-0.8em} &= \bar\psi(x)P_+\Big|_{x_0=0}\hspace*{-0.8em} =0 \,,\\
      P_-\psi(x)\Big|_{x_0=T}\hspace*{-0.8em} &= \rho'({\mathbf{x}})            \,,&
  \bar\psi(x)P_+\Big|_{x_0=T}\hspace*{-0.8em} &= \bar\rho'({\mathbf{x}})        \,,&
      P_+\psi(x)\Big|_{x_0=T}\hspace*{-0.8em} &= \bar\psi(x)P_-\Big|_{x_0=T}\hspace*{-0.8em} =0 \,,
\end{align}
with $P_\pm=\frac{1}{2}(1\pm\gamma_0)$.  Gauge fields are periodic in spatial
directions, whereas fermion fields are periodic up to global phases,
\begin{align}
    \psi(x+L\hat{\mathbf{k}}) &= e^{i\theta_k}\psi(x)       \,, & 
\bar\psi(x+L\hat{\mathbf{k}}) &= \bar\psi(x)e^{-i\theta_k}  \,.
\end{align}
The SF itself is the generating functional
\begin{align}
    \mathcal{Z}[C,\bar\rho,\rho;C',\bar\rho',\rho'] &=
    \int{\rm D}[U,\psi,\bar\psi]\,e^{-S[U,\bar\psi,\psi]} \,,
\end{align}
where the integral is performed over all fields with the specified boundary
values.  Expectation values of any product $\mathcal{O}$ of fields are then
given by
\begin{align}
\langle\mathcal{O}\rangle &= 
    \left\{\frac{1}{\mathcal{Z}}\int{\rm D}[U,\psi,\bar\psi]
    \,\mathcal{O} e^{-S[U,\bar\psi,\psi]}\right\}_{\bar\rho=\rho=\bar\rho'=\rho'=0} \,,
\end{align}
where $\mathcal{O}$ can involve, in particular, the ``boundary fields''
\begin{align}
    \zeta({\mathbf{x}})  &= \frac{\delta}{\delta\bar\rho({\mathbf{x}})}     \,, & 
\bar\zeta({\mathbf{x}})  &= -\frac{\delta}{\delta\rho({\mathbf{x}})} \,, & 
    \zeta'({\mathbf{x}}) &= \frac{\delta}{\delta\bar\rho'({\mathbf{x}})}    \,, & 
\bar\zeta'({\mathbf{x}}) &= -\frac{\delta}{\delta\rho'({\mathbf{x}})}\,.
\end{align}
The Dirichlet boundary conditions provide an infrared cutoff to the possible
wavelengths of quark and gluon fields, which allows one to study the theory
through simulations at vanishing quark mass.  The presence of non-trivial
boundary conditions requires, in general, additional counterterms to
renormalise the theory~\cite{Symanzik:1981wd,Luscher:1985iu,Luscher:1992an}.
In the case of the SF, it has been shown in Ref.~\cite{Sint:1995rb} that no
additional counterterms are needed with respect to the periodic case, except
for one boundary term that amounts to rescaling the boundary values of quark
fields by a logarithmically divergent factor, which is furthermore absent if
$\bar\rho=\rho=\bar\rho'=\rho'=0$. It then follows that the SF is finite after
the usual QCD renormalisation.

\section{Symanzik improvement of $T_{\mu\nu}$}\label{sec:improvement}

It is well established that $\mathrm{O}(a)$ improvement coefficients
(as well as scale-independent renormalisation constants) in lattice QCD
with Wilson fermions can be non-perturbatively determined by imposing
chiral Ward identities, which are consequences of the invariance of the
integration measure in the QCD functional integral representation of
expectation values under infinitesimal iso-vector transformations, to hold
on the lattice up to next-to-leading-order cutoff effects.
For applications of this approach to three-flavour QCD regularised with
the same lattice action as studied here, but in channels other than
the tensor one, see, for instance,
Refs.~\cite{Bulava:2016ktf,Heitger:2020mkp,Heitger:2020zaq,Heitger:2021bmg}.

In case of the flavour non-singlet tensor currents, our starting point to
derive an expression fixing the improvement coefficient $\cT$
is the general continuum axial Ward identity in its integrated form
\begin{multline}\label{eq:tensor_intWI}
 \int_{\partial R}\mathrm{d}^3x \, \langle A_0^a(x) \mathcal{O}_{\mathrm{int}}^b(y) \mathcal{O}_{\mathrm{ext}}^c(z) \rangle -2m \int_R \mathrm{d}^4x\, \langle P^a(x) \mathcal{O}_\mathrm{int}^b(y) \mathcal{O}_\mathrm{ext}^c(z) \rangle\\ 
= - \langle [\delta_\mathrm{A}^a\mathcal{O}_\mathrm{int}^b(y)] \mathcal{O}_\mathrm{ext}^c(z)\rangle \,,
\end{multline}
where $A_0^a$ and $P^a$ denote the axial vector current and the
pseudoscalar density, respectively, which are defined as
\begin{align}
    A^a_{0}(x) &= \bar{\psi}(x)\gamma_0\gamma_5 T^a \psi(x) \,, &
    P^a(x)     &= \bar{\psi}(x)\gamma_5 T^a \psi(x) \,.
\end{align}
As before, $T^a$ are the anti-Hermitean generators of $\mathrm{SU}(\Nf)$ acting
in flavour space.  In Eq.~\eqref{eq:tensor_intWI}, the composite fields
$\mathcal{O}_{\mathrm{int}}$ ($\mathcal{O}_{\mathrm{ext}}$) stand for
polynomials in the basic field operators that are localised in the interior
(exterior) of a space-time region $R$ with smooth boundary $\partial R$, i.e.,
that only have support inside (outside) $R$.  Recalling that the iso-vector
axial rotations underlying this Ward identity imply the infinitesimal
variations of the quark fields to read
\begin{align}
    \delta_\mathrm{A}^a\psi(x)       &\approx \mathrm{i}\gamma_5 T^a\psi(x)        \,, &
    \delta_\mathrm{A}^a\bar{\psi}(x) &\approx \mathrm{i}\bar{\psi}(x)\gamma_5 T^a  \,,
\end{align}
the behaviour of the tensor currents associated with these variations is
worked out straightforwardly using Leibniz's rule (where our Lie algebra
conventions are as in Ref.~\cite[Appendix~A]{Luscher:1996sc}):
\begin{align}
    \delta_\mathrm{A}^aT_{\mu\nu}^b(x) 
        &= -\bar{\psi}(x)T^a\gamma_5\sigma_{\mu\nu}T^b\psi(x)-\bar{\psi}(x)\sigma_{\mu\nu}\gamma_5T^bT^a\psi(x) \nonumber\\
        &= -\bar{\psi}(x)\gamma_5\sigma_{\mu\nu}(T^aT^b+T^bT^a)\psi(x) \nonumber\\
        &= d^{abc}\tilde{T}_{\mu\nu}^c(x) \,,\quad a\neq b \,, \\
    \delta_\mathrm{A}^a\tilde{T}_{\mu\nu}^b(x)
        &=d^{abc}T_{\mu\nu}^c(x) \,,\quad a\neq b \,.
\label{eq:dual_tensor_variation}
\end{align}
Here we introduced the dual tensor currents $\tilde{T}_{\mu\nu}$ as
\begin{align}\label{eq:dual_tensor_current}
    \tilde{T}_{\mu\nu}^a(x) &\equiv
     \mathrm{i}\bar{\psi}(x)\gamma_5\sigma_{\mu\nu}T^a\psi(x) = 
     -\frac{\mathrm{i}}{2}\epsilon_{\mu\nu\rho\sigma}\bar{\psi}(x)\sigma_{\rho\sigma}T^a\psi(x) \,,
\end{align}
where the second equality follows from the property
$\gamma_5\sigma_{\mu\nu}=-\frac{1}{2}\epsilon_{\mu\nu\rho\sigma}\sigma_{\rho\sigma}$.

We now exploit the freedom of a suitable choice for the internal operator
$\mathcal{O}_\mathrm{int}$ in Eq.~\eqref{eq:tensor_intWI} to just set it to
the dual tensor current, viz.
\begin{multline}
    \int_{\partial R}\mathrm{d}^3x 
        \,\langle A_0^a(x) \tilde{T}_{\mu\nu}^b(y) \mathcal{O}_\mathrm{ext}^c(z) \rangle 
        -2m \int_R \mathrm{d}^4x\, \langle P^a(x) \tilde{T}_{\mu\nu}^b(y) \mathcal{O}_\mathrm{ext}^c(z) \rangle \\
    = -d^{abd}\langle T_{\mu\nu}^d(y) \mathcal{O}_\mathrm{ext}^c(z)\rangle \,,
\end{multline}
which has non-vanishing r.h.s. for $\Nf\ge 3$ only. 
Using Eq.~\eqref{eq:dual_tensor_variation} for $a\neq b$ (to avoid mixing with the 
flavour-singlet tensor current), and in addition assuming $\mu=0$ and $\nu=k$ for the 
Dirac indices, we obtain:
\begin{multline}
    \int_{\partial R}\mathrm{d}^3x 
        \,\langle A_0^a(x) \tilde{T}_{0k}^b(y) \mathcal{O}_\mathrm{ext}^c(z) \rangle 
        -2m \int_R \mathrm{d}^4x\, \langle P^a(x) \tilde{T}_{0k}^b(y) \mathcal{O}_\mathrm{ext}^c(z) \rangle \\
    = -d^{abd}\langle T_{0k}^d(y) \mathcal{O}_\mathrm{ext}^c(z)\rangle \,.
\end{multline}
Inserting Eq.~\eqref{eq:dual_tensor_current} and keeping in mind that the
chromomagnetic components of the tensor currents do not require improvement,
cf. Eq.~\eqref{eq:chromo_magnetic_improvement},
the $\mathrm{O}(a)$ version of this lattice Ward identity then turns into:
\begin{align}\label{eq:tensor_intWI_Oa}
\epsilon_{0kij}\ZA&\bigg(\int_{\partial R}\mathrm{d}^3x \,\langle \big(A_0^a\big)^\mathrm{I}(x) T_{ij}^b(y) \mathcal{O}_\mathrm{ext}^c(z) \rangle -2m \int_R \mathrm{d}^4x\, \langle P^a(x) T_{ij}^b(y) \mathcal{O}_\mathrm{ext}^c(z) \rangle\bigg) \nonumber \\
& =\;\; 2d^{abd}\Big(\langle T_{0k}^d(y) \mathcal{O}_\mathrm{ext}^c(z)\rangle+a\cT\langle \tilde{\partial}_0V_k^d(y) \mathcal{O}_\mathrm{ext}^c(z)\rangle\Big) \,+\, \mathrm{O}(a^2) \,.
\end{align}
Note that in the chiral limit, in which we work in practice, the
$\mathrm{O}(a)$ improved axial vector current
$\big(A_{\mu}^a\big)^\mathrm{I}(x)=A_{\mu}^a(x)+a\cA\tilde{\partial}_\mu
P^a(x)$ receives finite multiplicative renormalisation via the factor
$\ZA(g_0^2)$, while any renormalisation factors for the tensor
currents and the (not yet specified) external operator
$\mathcal{O}_\mathrm{ext}$ appear on both sides of
Eq.~\eqref{eq:tensor_intWI_Oa} and thus cancel out.

Expression \eqref{eq:tensor_intWI_Oa} relates expectation values involving
chromomagnetic components of the tensor current to an expectation value of
its chromoelectric components.
As only the latter requires improvement in our specific setup, we can
employ Eq.~(\ref{eq:tensor_intWI_Oa}) to determine $\cT(g_0^2)$
non-perturbatively.
Even though this Ward identity holds for any tensor component separately,
we will numerically evaluate it by explicitly summing over the spatial
components $k$. 

\subsection{Non-perturbative determination in the Schr\"odinger functional scheme}\label{sec:cT_SF}

Our non-perturbative computation of the tensor currents' improvement
coefficient through numerical simulations works with a lattice discretisation
of QCD obeying Schr\"odinger functional boundary conditions (i.e., periodic in
space and Dirichlet in time).
Thanks to the gap in the spectrum of the Dirac operator thus introduced,
we do simulate the theory in the very close vicinity of the chiral limit that
is realised as the (unitary) point of vanishing degenerate sea and valence
quark masses, $am=0$ in short.

For the external operator $\mathcal{O}^c_\mathrm{ext}$ we now pick parity-odd
Schr\"odinger functional boundary fields~\cite{Luscher:1996sc}
\begin{align}
    \mathcal{O}^c[\Gamma]          &= 
        a^6\sum_{\mathbf{u},\mathbf{v}}\bar{\zeta}(\mathbf{u})\Gamma T^c\zeta(\mathbf v) \,, &
    \mathcal{O}^{\prime c}[\Gamma] &= 
        a^6\sum_{\mathbf{u},\mathbf{v}}\bar{\zeta}^\prime(\mathbf{u})\Gamma T^c\zeta^\prime(\mathbf v) \,,
\end{align}
where we choose $\Gamma=\gamma_k$ such that Eq.~\eqref{eq:tensor_intWI_Oa}
becomes (up to $\mathrm{O}(a^2)$ effects)\footnote{%
Recall that renormalisation factors of the boundary quark fields,
as well as of the tensor currents, cancel between the two sides of this
equation.}
\begin{multline}\label{epsilonZAtimesetc}
\epsilon_{0kij}\ZA\bigg(\int_{\partial R}\mathrm{d}^3x \,\langle 
    \big(A_0^a(x)\big)^\mathrm{I} T_{ij}^b(y) \mathcal{O}^c[\gamma_k] \rangle 
        -2m \int_R \mathrm{d}^4x\, \langle P^a(x) T_{ij}^b(y) \mathcal{O}^c[\gamma_k] \rangle\bigg) \\
 =  2d^{abd}\Big(\langle T_{0k}^d(y) \mathcal{O}^c[\gamma_k]\rangle
    +a\cT\langle \tilde{\partial}_0V_k^d(y) \mathcal{O}^c[\gamma_k]\rangle\Big) \,.
\end{multline}
Another option for the Dirac structure would be to choose
$\mathcal{O}[\gamma_0\gamma_k]$ as the boundary interpolator.  However, because
of the Schr\"odinger functional boundary conditions, the projection operator
$P_\pm=\frac{1}{2}(1\pm\gamma_0)$ mixes $\gamma_k$ and $\gamma_0\gamma_k$,
which renders this choice ambiguous.

To express Eq.~\eqref{epsilonZAtimesetc} in terms of correlation functions
and thereby make it accessible to numerical calculation, we now define the
Schr\"odinger functional (boundary-to-bulk) correlator as
$\kAsig(x_0,y_0)=-\frac{1}{(\Nf^2-1)6}\,d^{abc}\kAsig^{abc}(x_0,y_0)$, where
\begin{align}
\kAsig^{abc}(x_0,y_0)=&\,-\frac{1}{2}\frac{a^6}{L^3}\sum_{\mathbf{x},\mathbf{y}}\epsilon_{0kij}\langle A_0^a(x) \bar{\psi}(y)\sigma_{ij}T^b\psi(y) \mathcal{O}^c[\gamma_k] \rangle \nonumber \\
=&\;\frac{1}{2}\frac{a^6}{L^3}\,\epsilon_{0kij}\epsilon_{ijl}\sum_{\mathbf{x},\mathbf{y}}\langle A_0^a(x) \bar{\psi}(y)
\begin{pmatrix} \sigma_l & 0 \\ 0 & \sigma_l \end{pmatrix}
T^b\psi(y) \mathcal{O}^c[\gamma_k] \rangle \nonumber \\
=&\;\frac{a^6}{L^3}\,\delta_{kl}\sum_{\mathbf{x},\mathbf{y}}\langle A_0^a(x) \bar{\psi}(y)
\begin{pmatrix} \sigma_l & 0 \\ 0 & \sigma_l \end{pmatrix}
T^b\psi(y) \mathcal{O}^c[\gamma_k] \rangle \nonumber \\
=&\;\frac{a^6}{L^3}\sum_{\mathbf{x},\mathbf{y}}\langle A_0^a(x) \bar{\psi}(y)
\begin{pmatrix} \sigma_k & 0 \\ 0 & \sigma_k \end{pmatrix}
T^b\psi(y) \mathcal{O}^c[\gamma_k] \rangle \,,
\end{align}
using $\sigma_{ij}= -\epsilon_{ijl} \begin{pmatrix} \sigma_l & 0 \\ 0 &
\sigma_l \end{pmatrix}$ in the first step.
The analogous correlator $\kPsig$ in the pseudoscalar channel only differs from
$\kAsig$ by the fact that in $\kAsig^{abc}(x_0,y_0)$ the temporal axial vector
current $A_0^a(x)$ is replaced with the pseudoscalar density $P^a(x)$.
Altogether this finally leads to
\begin{multline}
 \ZA\Big(\kAsig^{abc}(x_0,y_0)+a\cA\tilde{\partial}_0\kPsig^{abc}(x_0,y_0)-2m\ktilPsig^{abc}(x_0,y_0)\Big) \\
= \mathrm{i}d^{abd}\Big( \kT^{dc}(y_0)+a\cT\tilde{\partial}_0\kV^{dc}(y_0) \Big) \,,
\end{multline}
which can be solved for $\cT$:
\begin{align}\label{eq:orginal_impr_con}
 a\cT &= \frac{\ZA\Big(\kAsig^{abc}(x_0,y_0)+a\cA\tilde{\partial}_0\kPsig^{abc}(x_0,y_0)
    -2m\ktilPsig^{abc}(x_0,y_0)\Big)-\mathrm{i}d^{abd}\kT^{dc}(y_0)}%
    {\mathrm{i}d^{abd}\tilde{\partial}_0\kV^{dc}(y_0)} \,,
\end{align}
up to $\mathrm{O}(a^2)$ corrections.
The definitions of the boundary-to-bulk correlators $\kT$ and $\kV$ are
\begin{align}
\label{eq:kT}
 \kT^{ab}(x_0) &= -\frac{1}{6}\frac{a^3}{L^3}\sum_{\mathbf{x}}\langle T_{0k}^a(x_0,\mathbf{x})\mathcal{O}^b[\gamma_k] \rangle \,, \\
\label{eq:kV}
 \kV^{ab}(x_0) &= -\frac{1}{6}\frac{a^3}{L^3}\sum_{\mathbf{x}}\langle V_k^a(x_0,\mathbf{x})\mathcal{O}^b[\gamma_k] \rangle \,,
\end{align}
with $V_k^a(x)=\bar\psi(x)\gamma_k T^a\psi(x)$.
The correlation functions $\ktilPsig^{abc}(x_0,y_0)$ are
defined in the same way as $\kPsig^{abc}(x_0,y_0)$, but incorporating
a proper weight approximating the integration via the trapezoidal rule 
(cf.~\cite[Eq.~(B.9)]{Heitger:2020zaq}) when translating
Eq.~(\ref{epsilonZAtimesetc}) to the lattice.
Also note that for the moment we have suppressed the implicit dependence of
the l.h.s. of this equation on the timeslice arguments $x_0$ and $y_0$;
our specific choices for its numerical evaluation will be detailed later.
As external inputs for the computation of $\cT$,
the improvement coefficient $\cA$~\cite{Bulava:2015bxa} and the
renormalisation factor $\ZA$ of the axial vector current are required,
for which the non-perturbative three-flavour QCD determinations with the
same lattice action (as functions of $g_0^2$) of
Refs.~\cite{Bulava:2016ktf,DallaBrida:2018tpn} are available; the
renormalisation factor is based on the chirally rotated Schr\"odinger
functional~\cite{DallaBrida:2018tpn} and from now on referred to as $\ZA^\chi$.

Within a similar computation for $\mathrm{O}(a)$ improvement of the vector
current~\cite{Heitger:2020zaq} it was found that the very precise results on
$\ZA^\chi$ appear to have a significant lattice spacing ambiguity of
$\mathrm{O}(a^3)$, in addition to the leading $\mathrm{O}(a^2)$ one.
Whereas this is immaterial when one is interested in matrix elements of the
axial vector current, it can have a non-negligible impact on the determination
of~$\cT$.
The reason lies in the fact that $\cT$ obtained from the identity
derived above arises as the difference of two terms, which are orders of
magnitude larger than their difference and where only one of the two is
multiplied by $\ZA^\chi$.
Therefore, a small change in $\ZA^\chi$ can propagate into a significant
change in $\cT$.
While from the Symanzik improvement programme point of view this is not
worrisome---since the ambiguity in $\ZA^\chi$ is beyond the order
in $a$ we are interested in for $\cT$---its absolute magnitude can still
be sizable~\cite{Heitger:2020zaq}.
In order to tame the potential influence of these higher-order effects, 
we hence propose an alternative improvement condition for $\cT$, 
which instead of $\ZA^\chi$ takes the ratio $\ZA^\chi/\ZV^\chi$ as an input,
with $\ZV^\chi$ the vector current's renormalisation factor also
known from the chirally rotated Schr\"odinger functional
study~\cite{DallaBrida:2018tpn}.
As a consequence, the $\mathrm{O}(a^2)$ and $\mathrm{O}(a^3)$ ambiguities
will cancel numerically in this ratio when using the parametrisations of
$\ZA^\chi$ and $\ZV^\chi$ in terms of $g_0^2$ from Ref.~\cite{DallaBrida:2018tpn}.
To reformulate Eq.~\eqref{eq:orginal_impr_con} so that it incorporates
$\ZA^\chi/\ZV^\chi$, we make use of the vector Ward identity
which, when implemented with Schr\"odinger functional boundary fields
(see, e.g., Ref.~\cite{Heitger:2020zaq}), yields an independent
determination of the vector current renormalisation constant $\ZV$,
\begin{align}
\label{eq:def_ZV}
\frac{f_1}{\FV(x_0)}=\ZV+\mathrm{O}(a^2) \,,
\end{align}
with appropriate three-point and boundary-to-boundary Schr\"odinger functional
correlation functions
\begin{align}
 \FV(x_0) 
     &= -\frac{a^3}{2L^6}\rmi\epsilon^{abc}\sum_{\mathbf{x}}\big\langle \mathcal{O}^{\prime a}[\gamma_5]V_0^b(x_0,\mathbf{x})\mathcal{O}^c[\gamma_5]\big\rangle \,,\\
\label{eq:f1}
 f_1 &= -\frac{1}{2L^6}\big\langle \mathcal{O}^{\prime a}[\gamma_5]\mathcal{O}^a[\gamma_5]\big\rangle \,.
\end{align}
In practice, a renormalisation condition such as~\eqref{eq:def_ZV} is always
understood to be numerically evaluated for every timeslice and averaged over
a certain plateau range, which completes the definition of $\ZV$.
Then, in the $\mathrm{O}(a)$ improved theory, the ratio of $\ZV$
fixed in this fashion and $\ZV^\chi$ of Ref.~\cite{DallaBrida:2018tpn}
is just unity up to $\mathrm{O}(a^2)$ cutoff effects:
\begin{align}
\frac{\ZV}{\ZV^\chi}=\frac{f_1}{\ZV^\chi \FV(x_0)}=1+\mathrm{O}(a^2) \,.
\end{align}
To this order we thus can extend the prefactor of the first term in the
numerator of Eq.~\eqref{eq:orginal_impr_con} by $1=f_1/(\ZV^\chi\FV)$ to arrive
at an alternative improvement condition for $\cT$,
\begin{align}\label{eq:new_impr_con}
    a\cTalt &= \frac{\frac{\ZA^\chi}{\ZV^\chi}\frac{f_1}{\FV(x_0)}\Big(\kAsig^{abc}(x_0,y_0)+a\cA\tilde{\partial}_0\kPsig^{abc}(x_0,y_0)-2m\ktilPsig^{abc}(x_0,y_0)\big)-\mathrm{i}d^{abd}\kT^{dc}(y_0)}{\mathrm{i}d^{abd}\tilde{\partial}_0\kV^{dc}(y_0)} \,,
\end{align}
in which non-perturbative values on $\cA$ from Ref.~\cite{Bulava:2015bxa} 
and the ratio $\ZA^\chi/\ZV^\chi$ from Ref.~\cite{DallaBrida:2018tpn}
are to be inserted as external inputs.

\subsection{Analysis and results}\label{sec:cT_SF_results}

Our computations and analyses are based on gauge field ensembles
generated by numerical simulations of mass-degenerate three-flavour lattice
QCD with non-perturbatively $\mathrm{O}(a)$ improved Wilson quarks and
tree-level Symanzik-improved gluons for earlier determinations
of improvement coefficients and renormalisation
factors~\cite{Bulava:2015bxa,Bulava:2016ktf,deDivitiis:2019xla}.
They satisfy Schr\"odinger functional boundary conditions and describe a 
line of constant physics (LCP) defined by a fixed physical $L^3\times T$ 
box of size $L\approx 1.2\,\mathrm{fm}\approx 2T/3$;
its parameters are listed in \Cref{tab:gauge_parameters} of
Appendix~\ref{app:improvement}.
The lattice spacings of these ensembles lie within
$0.042\,\fm\lesssim a\lesssim 0.105\,\fm$ and suitably match
the range of lattice spacings that is typically accessible in large-volume
simulations, thereby making our results for $\cT$ useful for physics
applications.  
For all further technical and algorithmic details we refer to the
aforementioned references; we only point out that the known issue of topology 
freezing towards the continuum limit is accounted for in a theoretically 
sound way by projecting all expectation values to the topologically trivial 
sector by reweighting, and that for the error analysis we use the 
$\Gamma$-method~\cite{Wolff:2003sm}  in the Python implementation described
in Ref.~\cite{Joswig:2022qfe}, which includes effects of critical slowing
down along Ref.~\cite{Schaefer:2010hu} and automatic differentiation
techniques as suggested in Ref.~\cite{Ramos:2018vgu}.

\begin{figure}[t]
	\centering
	\includegraphics[width=0.9\linewidth]{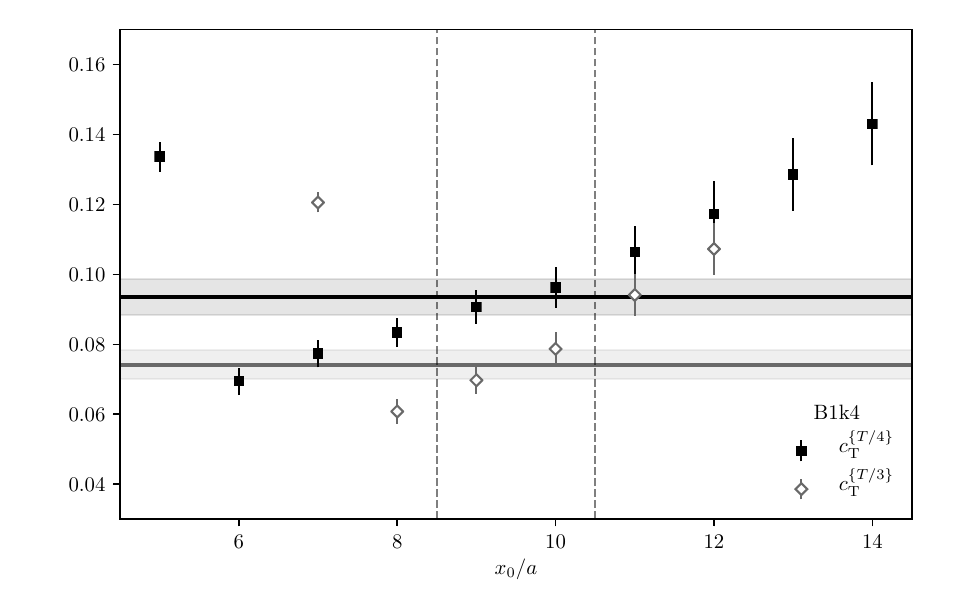}
	\caption{Euclidean time dependence of the two determinations of $\cT$
                 for ensemble B1k4 ($\beta=3.512$). The dashed vertical lines
                 enclose the plateau region, the horizontal lines with shaded
                 bands correspond to the plateau values and their respective
		 uncertainties.
	}
	\label{fig:cT_plateau}
\end{figure}
The lattice setup to extract $\cT$ realised here ensures that 
the initial chiral Ward identity, and thereby also the improvement condition
formulae~\eqref{eq:orginal_impr_con} and~\eqref{eq:new_impr_con} implied
by it, are imposed on an LCP, along which all physical length scales in
correlation functions are kept constant and only the lattice spacing $a$
changes when the bare coupling $g_0$ is varied.
According to Symanzik's local effective theory of cutoff effects,
the resulting estimates of improvement coefficients such as $\cT$
are then supposed to exhibit a smooth dependence on $g_0^2=6/\beta$.
As a consequence, any remaining ambiguities in them that could emerge through
another choice of LCP or from a different improvement condition will
asymptotically disappear towards the continuum limit ($a\to 0$) at a minimum
rate $\propto a$.
We thus expect our final results for $\cT(g_0^2)$ to be potentially
affected by $\mathrm{O}(a)$ corrections only; however, in view of
Eq.~\eqref{eq:improved_tensor_current}, the latter are beyond the order
one is sensitive to in the $\mathrm{O}(a)$ improved theory.
Equivalently, any effects of these unavoidable $\mathrm{O}(a)$ ambiguities
in hadronic matrix elements or other quantities involving improved
(and renormalised) tensor currents will extrapolate to zero in the continuum
limit.

The details of our analysis closely follow the similar investigation
\cite{Heitger:2020zaq} of $\mathrm{O}(a)$ improvement in the vector channel.
Following our experiences there, 
Eqs.~\eqref{eq:orginal_impr_con} and~\eqref{eq:new_impr_con} are evaluated
with two different operator positions $t_1=T-t_2 \in \lbrace T/4,T/3 \rbrace$
and averaged over the two central values of $x_0$,
where $\cA(g_0^2)$ is taken from Ref.~\cite{Bulava:2015bxa}, while for the
axial and vector currents' normalisation constants we employ $\ZA^\chi$ and
$\ZV^\chi$ calculated within the chirally rotated Schr\"odinger
functional~\cite{DallaBrida:2018tpn}, as explained in Section~\ref{sec:cT_SF}.
The $x_0$-dependence of the improvement coefficient on an exemplary
ensemble is displayed in Figure~\ref{fig:cT_plateau}.
Although the local estimator as a function of $x_0$ does not develop a clear
plateau, as was also observed in similar studies
\cite{Gerardin:2018kpy,Heitger:2020zaq}, our Ward-identity approach
(formulated on the operator level) in conjunction with the LCP setup
guarantees that any choice of $x_0$ and $t_1$ within the $x_0$-region
furthest from the boundaries is allowed, as long as all physical length
scales are kept constant among the different ensembles.

\begin{figure}[t]
	\centering
	\includegraphics[width=0.9\linewidth]{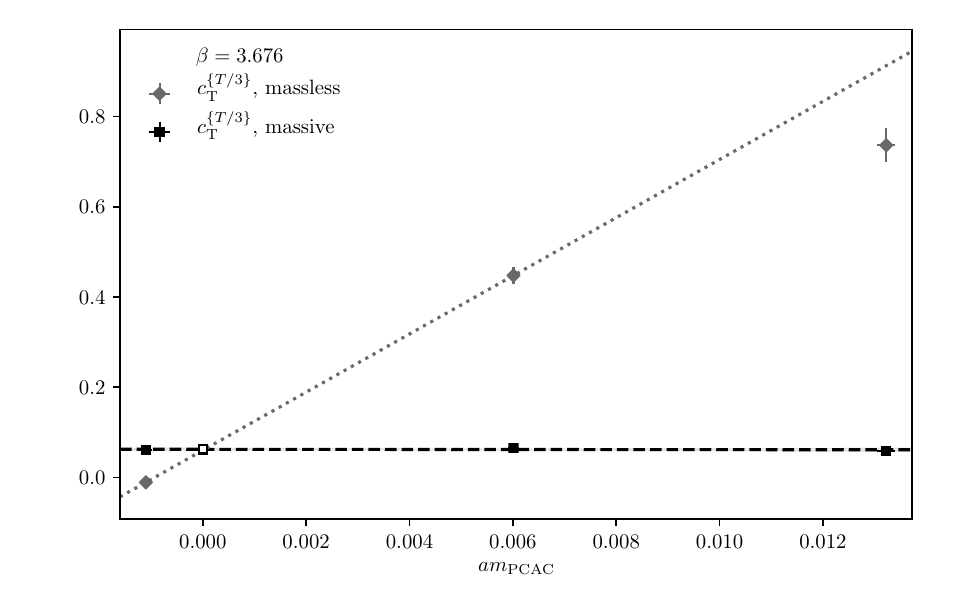}
    \caption{Example of a chiral extrapolation of $\cT$, here for 
    	     $\cT^{\Xstyle\lbrace T/3\rbrace}$ at $\beta=3.676$, with and without 
    	     the mass term in the Ward identity~\eqref{eq:orginal_impr_con}. 
    	     The dashed lines indicate linear fits to the data, with shaded bands 
    	     corresponding to the uncertainties of the fits.
    	     The open symbol marks the value in the chiral limit.
            }
	\label{fig:cT_chiral_extrapolation}
\end{figure}
Figure~\ref{fig:cT_chiral_extrapolation} shows an exemplary extrapolation of 
$\cT$ to the chiral limit ($am_{\rm PCAC}\to 0$) for the case of the improvement 
condition yielding $\cT^{\Xstyle\lbrace T/3\rbrace}$ (i.e., the value of $\cT$
obtained when the operator is inserted at $t_1=T/3$), once with the explicit 
mass term (see Eq.~\eqref{eq:orginal_impr_con}) and once without it.
Note that our quark mass definition, including the chiral limit as the
point of zero sea (and valence) quark masses, is the one based on the 
partially conserved axial current (PCAC) relation; see Eq.~\eqref{eq:pcac_mass} 
in Section~\ref{sec:RG_uSF} for its explicit lattice prescription.
The mass dependence of the data in Figure~\ref{fig:cT_chiral_extrapolation} 
illustrates very distinctly that accounting for the mass term in the evaluation 
of the Ward identity entails an almost flat quark mass dependence and hence 
a very stable extrapolation.
The chirally extrapolated results for the considered variants to extract
$\cT$ are compiled for all ensembles
in Table~\ref{tab:cT_results} of Appendix~\ref{app:improvement}.

In Figure~\ref{fig:cT_overview} we compare the different sets of results of
$\cT$ from this work with the one-loop prediction for 
$\cT$ of Ref.~\cite{Taniguchi:1998pf} for the tree-level Symanzik-improved 
(namely, the L\"uscher--Weisz) gauge action.
\begin{figure}[t]
	\centering
	\includegraphics[width=0.9\linewidth]{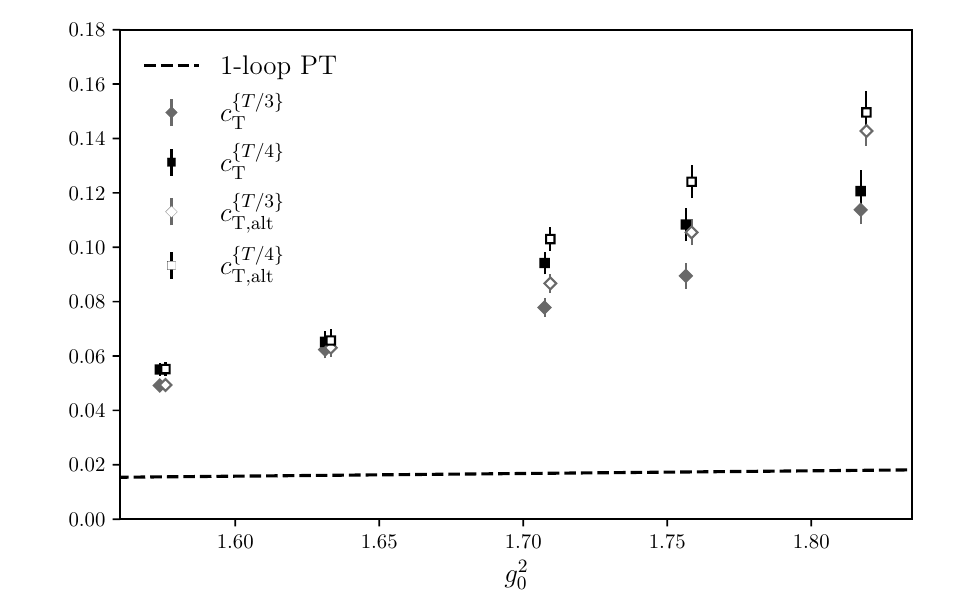}
	\caption{Different determinations of $\cT$ as a function of the squared 
	         bare coupling, $g_0^2$, in comparison to one-loop perturbation
                 theory~\cite{Taniguchi:1998pf}.}
	\label{fig:cT_overview}
\end{figure}
All of our non-perturbative determinations strongly deviate from perturbation
theory and approach the one-loop prediction only as $g_0\to 0$.
In particular, these four variants agree within errors for the two smallest
lattice spacings, while they exhibit larger (albeit monotonic) spreads for
the coarser ones.
This behaviour reinforces that, as argued above, the intrinsic
$\mathrm{O}(a)$ ambiguities between determinations based on different
improvement conditions vanish smoothly towards the continuum limit, i.e.,
for $\beta=6/g_0^2\to\infty$.

\begin{figure}[t]
	\centering
	\includegraphics[width=0.9\linewidth]{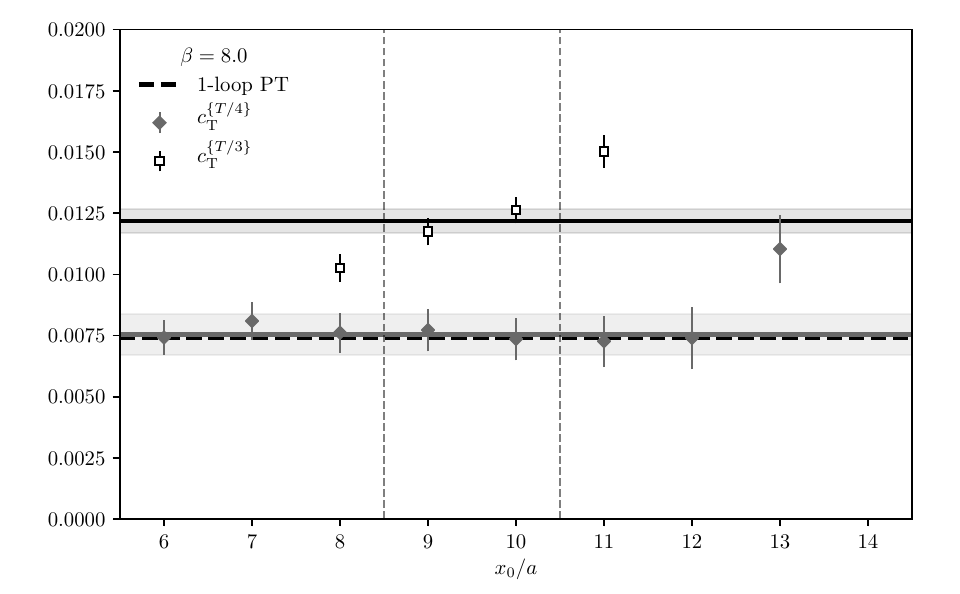}
	\caption{Timeslice dependence of $\cT$ at $\beta=8$ compared to
                 the one-loop perturbation theory prediction of
                 Ref.~\cite{Taniguchi:1998pf} (dashed horizontal line).
                 Solid horizontal lines and surrounding 
		 shaded areas display the plateau averages and errors of the
                 two non-perturbative estimators with $t_1=T/4$ and $t_1=T/3$,
                 respectively.
		 The corresponding current quark mass vanishes within error.}
	\label{fig:cT_weak_coupling}
	\end{figure}

In order to settle on a final one among the different, equally admissible
determinations labelled as $\cT^{\Xstyle\lbrace T/4,T/3 \rbrace}$ and 
$\cTalt^{\Xstyle\lbrace T/4,T/3 \rbrace}$, we consulted the behaviour in 
the region of weaker couplings: this is motivated by the fact that $\cT$ is
also to be applied in the present context of non-perturbatively solving the
scale-dependent renormalisation problem of the tensor current via
step-scaling methods in Section~\ref{sec:renormalisation}.
For this purpose of evaluating the renormalisation factor $\ZT(g_0^2,a\mu)$,
we specifically need the improvement coefficient at bare couplings 
$1\lesssim g_0^2\lesssim 1.7$ considerably smaller than the ones covered
by the data discussed so far.
Therefore, we have performed an additional simulation at $\beta=8$
and volume $(L^3\times T)/a^4=16^3\times 24$ (allowing relaxation of the 
LCP condition in this almost perturbative regime), which is to be included
as a further constraint in a later interpolation formula for $\cT(g_0^2)$.
In Figure~\ref{fig:cT_weak_coupling} we illustrate the resulting timeslice 
dependence of  the (standard) improvement condition in comparison to the 
one-loop perturbative prediction~\cite{Taniguchi:1998pf}.
While the general picture is very similar to what is encountered in the
stronger coupling region, an important observation is that the
determination with operator insertion point $t_1=T/4$ nicely agrees with
the perturbative prediction, whereas $t_1=T/3$ shows a significant deviation.
Although this is not in disagreement with the theoretical expectation
(rather, it must be seen as an $\mathrm{O}(a)$ ambiguity), we eventually 
chose $\cTalt^{\Xstyle\lbrace T/4 \rbrace}$ as our preferred estimator of 
$\cT$, since it exhibits closer agreement with perturbation theory in the 
weak-coupling regime.

\begin{figure}[t]
	\centering
	\includegraphics[width=0.9\linewidth]{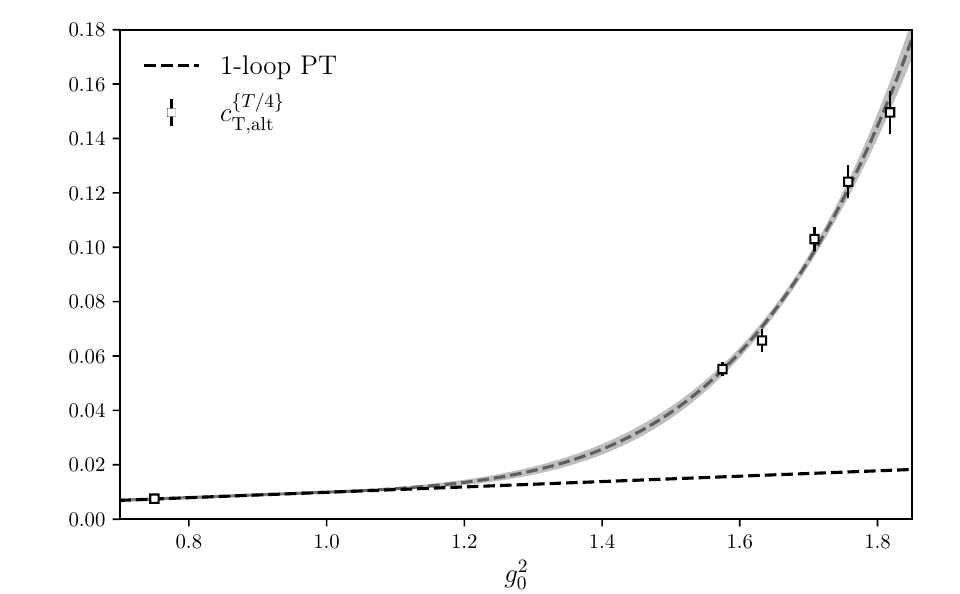}
	\caption{Interpolation formula~\eqref{eq:interpolation_cT} of our
                 preferred determination of $\cT$ in comparison to one-loop
                 perturbation theory~\cite{Taniguchi:1998pf}.
	}
	\label{fig:cT_interpolation}
\end{figure}

Figure~\ref{fig:cT_interpolation} presents the results of our
preferred determination of the improvement coefficient of the tensor
currents, together with the perturbative prediction and the continuous
interpolation of these final non-perturbative results in terms of $g_0^2$.
The fit formula is inspired by the leading term in the perturbative relation
between the lattice spacing and the $\beta$-function, $b_0=9/(4\pi)^2$ 
being the corresponding universal coefficient for $\Nf=3$, constrained by 
one-loop perturbation theory~\cite{Taniguchi:1998pf} as $g_0^2\rightarrow 0$.
The $g_0^2$-dependence in the range covered by the data is best represented
by a parametrisation of the form
\begin{subequations}
	\label{eq:interpolation_cT}
	\begin{align}
	    \cT(g_0^2)=0.00741\CF g_0^2\,\Big[1+\exp\big\lbrace-1/(2b_0g_0^2)
            \big\rbrace\big(p_1+p_2 g_0^2\big)\Big]\,,
	\end{align}
	with
	\begin{align}
            (p_i) &=
		\begin{pmatrix}
        		-1.232 \\
        		+1.203
    \end{pmatrix}\cdot 10^{3}   \,, &
     \operatorname{cov}(p_i,p_j)=
		\begin{pmatrix}
				+17.3880 & -10.2809 \\
				-10.2809 & +6.09472 \\
        \end{pmatrix}\cdot 10^{4}\,.
	\end{align}
\end{subequations}
It describes the five data points at stronger couplings and the weak-coupling
data point ($\beta=8$) with $\chi^2/{\rm d.o.f.}=0.851$.  Let us stress again
that this parametrisation and the one-loop behaviour agree almost perfectly up
to bare couplings of $g_0^2\approx 1$, whereas at larger ones a significant
departure from perturbation theory is clearly visible.

The interpolation formula \eqref{eq:interpolation_cT} holds for three-flavour
lattice QCD with $\mathrm{O}(a)$ improved Wilson fermions and a tree-level
Symanzik-improved gauge action, as also partly employed in the present work.
In particular, the bare lattice couplings used here largely overlap with those
of the $\Nf=2+1$ QCD gauge field configuration ensembles by the CLS
collaboration~\cite{Bruno:2014jqa,Bruno:2016plf,Bali:2016umi,Mohler:2017wnb,RQCD:2022xux},
which were generated with exactly this discretisation and provide a broad
landscape of pion masses and lattice spacings, aimed to various
phenomenological lattice QCD applications.
In order to make our results usable within future computations on the CLS
ensembles involving $\mathrm{O}(a)$ improved tensor currents, we finally also
interpolate (for $\beta=3.85$: slightly extrapolate) our results to the CLS
values of $\beta$.  These estimates for $\cT$ are collected in
Table~\ref{tab:cT_cls}.
\begin{table}[t]
    \centering
    \begin{tabular}{llllllll}
	\toprule
	$\beta$ & 3.34     & 3.4      & 3.46     & 3.55     & 3.7      & 3.85      \\\midrule
    $\cT$   & 0.143(5) & 0.125(3) & 0.110(3) & 0.091(2) & 0.067(2) & 0.051(3)  \\
	\bottomrule
    \end{tabular}
    \caption{$\cT$-results for the inverse gauge couplings $\beta$
             employed in the $\Nf=2+1$ CLS
             calculations~\cite{Bruno:2014jqa,Bali:2016umi,Mohler:2017wnb}. 
             The errors are the statistical uncertainties propagated from the
             interpolation formula \eqref{eq:interpolation_cT}, except for 
             $\beta=3.85$, which is slightly outside of the coupling range
             covered by our data. We thus add 50\% of the size of the
             statistical error in quadrature as a systematic uncertainty to
             account for this.
            }
    \label{tab:cT_cls}
\end{table}
%

\section{Renormalisation of $T_{\mu\nu}$}\label{sec:renormalisation}

\subsection{Renormalisation schemes and strategy}

Next, we discuss the renormalisation of the tensor current. Our strategy
follows the standard non-perturbative renormalisation and running setup by the
ALPHA collaboration, in particular in the context of $\Nf=3$ QCD.  We define
the RGI tensor current following Eq.~\eqref{rgi_operator},
\begin{gather}
\hat{T}_{\mu\nu} = \overline{T}_{\mu\nu}(\mu) \left[\frac{\overline{g}^2(\mu)}{4\pi}\right]^{-\gamT^{(0)}/2b_0}
\exp\left\{
-\int_0^{\overline{g}(\mu)}{\rm d}g\left[\frac{\gamT(g)}{\beta(g)}-\frac{\gamT^{(0)}}{b_0g}\right]
\right\}\,,\label{eq:RGI_tensor}
\end{gather}
where $\overline{T}_{\mu\nu}(\mu)$ is the renormalised tensor current in the
continuum, $\overline{g}(\mu)$ is some renormalised coupling, $\beta$ and
$\gamT$ are the $\beta$-function and the tensor anomalous dimension,
respectively, and $b_0,\gamT^{(0)}$ their leading perturbative coefficients.
We shall employ two different mass-independent, finite-volume renormalisation
schemes, defined by the renormalisation conditions
\begin{align}
  \ZT^{\ff}(g_0^2,a/L)\cdot \frac{\kT^\mathrm{I}(T/2)}{\sqrt{f_1}} &=
        \frac{\kT^\mathrm{I}(T/2)}{\sqrt{f_1}}\Bigg|_{\mbox{\tiny{tree-level}}} \,,\\
  \ZT^{\kk}(g_0^2,a/L)\cdot \frac{\kT^\mathrm{I}(T/2)}{\sqrt{k_1}} &=
        \frac{\kT^\mathrm{I}(T/2)}{\sqrt{k_1}}\Bigg|_{\mbox{\tiny{tree-level}}} \,,
\end{align}
where
\begin{gather}
    \kT^\mathrm{I}(x_0) = \kT(x_0) + a \cT\partial_0\kV(x_0)\,,
\end{gather}
and the correlation functions $\kT \equiv \frac{1}{\Nf^2-1}\kT^{aa}$ and $\kV
\equiv\frac{1}{\Nf^2-1} \kV^{aa}$ have been introduced in Eqs.~\eqref{eq:kT}
and~\eqref{eq:kV}. 
The boundary-to-boundary correlator $k_1$, similar to $f_1$ introduced in
Eq.~\eqref{eq:f1}, is given by
\begin{gather}
        k_1      = -\frac{1}{6L^6}\langle \mathcal{O}^{\prime a}[\gamma_k]\mathcal{O}^{a}[\gamma_k] \rangle \,.
\end{gather}

Having at our disposal these non-perturbative renormalisation constants serves
two purposes: we can renormalise the tensor current at any given scale
$\mu=1/L$ through Eq.~\eqref{eq:renormalized_current}, then trace the
renormalisation-group evolution of the current by introducing the step-scaling
functions
\begin{align}
  \sigT(u) &\equiv \lim_{a\to 0} \SigT(u,a/L)
            \equiv \lim_{a\to 0} \left.\frac{\ZT(g_0^2,a/(2L))}{\ZT(g_0^2,a/L)}\right|_{u=\overline{g}^2(1/L)}
                  \hspace*{-1em}= \exp\left\{ \int_{\gbar(L^{-1})}^{\gbar((2L)^{-1})}\kern-10pt \mathrm{d}g
                   \frac{\gamT(g)}{\beta(g)}\right\} \,.
\end{align}
By computing $\SigT^{\ff}(u,a/L)$ and $\SigT^{\kk}(u,a/L)$ at several values of
$u$ and $a/L$ it is possible to obtain the respective continuum versions
$\sigT^{\ff/\kk}(u)$, and hence $\gamT^{\ff/\kk}$, non-perturbatively for a
wide range of scales.
Recall that the step-scaling function is a particular case of the general
solution of the RGE~\eqref{oper_RGE} in terms of an RG evolution operator
$U(\mu_2,\mu_1)$, viz.
\begin{gather}
\label{eq:RGevol}
\overline{\mathcal{O}}(\mu_2) = U(\mu_2,\mu_1)\overline{\mathcal{O}}(\mu_1)\,; \qquad
U(\mu_2,\mu_1) = \exp\left\{ \int_{\gbar(\mu_1)}^{\gbar(\mu_2)}\kern-5pt \mathrm{d}g \frac{\gamma_\mathcal{O}(g)}{\beta(g)}\right\}\,,
\end{gather}
so that $\sigT(u)=U(\gbar((2L)^{-1}),\gbar(L^{-1}))$ with $u=\gbar(L^{-1})$ and
the appropriate anomalous dimension $\gamT$ of the tensor operator used in
Eq.~\eqref{eq:RGevol}.
In order to determine the anomalous dimension, we shall follow the same
strategy as for quark masses~\cite{Campos:2018ahf}. Using the notation in
Eqs.~\eqref{oper_RGE} and~\eqref{eq:RGevol}, we factorise
Eq.~\eqref{rgi_operator} as
\begin{gather}\label{eq:master}
\begin{split}
\hat{T}_{\mu\nu} &= \hat c(\muhad)\overline{T}_{\mu\nu}(\muhad)\\
&=
\underbrace{\hat c(\mu_{\rm\scriptscriptstyle PT})}_\text{PT}\,
\underbrace{U(\mu_{\rm\scriptscriptstyle PT},\mu_0/2)}_\text{SF}\,
\underbrace{U(\mu_0/2,\muhad)}_\text{GF}\,
\overline{T}_{\mu\nu}(\muhad)\,,
\end{split}
\end{gather}
where $\muhad$ is a low-energy scale of the order of
$\Lambda_{\rm\scriptscriptstyle QCD}$, $\mu_{\rm\scriptscriptstyle PT}$ is some
high-energy scale, of the order of the electroweak scale, where perturbation
theory is safe (next-to-leading-order predictions are available in our case),
$\mu_0$ is an intermediate scale of the order of $4\,\GeV$, and the factors
labelled ``GF'' and ``SF'' are computed using gradient-flow and SF
non-perturbative couplings, respectively (see Ref.~\cite{Campos:2018ahf} for a
detailed explanation, a full reference list, and any unexplained notation).
The key points in the whole setup are that each of these factors, except for
the first one, can be computed non-perturbatively and taken to the continuum
limit with fully controlled systematics, and that the connection to the RGI
allows one to match the result to any other renormalisation scheme convenient
for phenomenology.

\subsection{RG running at high energies}\label{sec:RG_uSF}

In the high-energy regime we have performed simulations at eight values of the
renormalised SF coupling
\begin{align}
    \uu[SF] &= \{1.1100, 1.1844, 1.2656, 1.3627, 1.4808, 1.6173, 1.7943, 2.0120\} \,,
\end{align}
using the Wilson plaquette action~\cite{Wilson:1974sk} and an
$\mathrm{O}(a)$-improved Wilson fermion action~\cite{Luscher:1996sc}, with the
non-perturbative value for $c_\mathrm{sw}$ from Ref.~\cite{JLQCD:2004vmw} and
one-loop~\cite{Luscher:1996vw} and two-loop~\cite{Bode:1999sm} values for the
boundary improvement coefficients $\cttil$ and $\ct$, respectively. Simulations
are performed at three different values of the (inverse) lattice spacing
$L/a=6,8,12$, except for the strongest coupling $\uu[SF]=2.012$ where a fourth,
finer lattice spacing $L/a=16$ is used ($L$ is implicitly fixed through $u$).
The three quarks in all simulations are tuned to be massless by demanding that
the PCAC mass,
\begin{align}
	m(g_{0}^{2},\kappa)=\left.\frac{{1\over 2}(%
		\partial_{0}^{*}+\partial_{0}^{\phantom{*}})f_{\rm A}(x_{0})+a \cA \partial_{0}^{*}\partial_{0}^{\phantom{*}}f_{\rm\scriptscriptstyle P}(%
		x_{0})}{2f_{\rm P}(x_{0})}\right|_{x_{0}=T/2}\,,\label{eq:pcac_mass}
\end{align}
vanishes, using the improvement coefficient $\cA(g_0^2)=-0.005680(2)\CF g_0^2$
to one-loop order in perturbation theory~\cite{Luscher:1996vw,Sint:1997jx}.
All simulations were performed with a variant of the \texttt{openQCD}
code~\cite{openQCD}.

Since our computation of the improvement coefficient $\cT$ is available only
for tree-level Symanzik-improved gauge action, we use its one-loop value for
the plaquette gauge action, determined in Ref.~\cite{Sint:1997jx}. Note that
this is not expected to have a major impact, since the lattices employed in the
high-energy region are very close to the continuum limit and the residual
$O(ag_0^4)$ cutoff effects should be highly suppressed.  Furtheremore, the
step-scaling functions $\SigT$ obtained from these simulations are corrected by
subtracting the cutoff effects to all orders in $a$ and leading order in
$g_0^2$, as described in Ref.~\cite[Sec.~4.2]{Pena:2017hct}.

\subsubsection{Determination of the anomalous dimension}

\begin{table}[t]
    \small
	\centering
	\renewcommand{\arraystretch}{1.25}
	\setlength{\tabcolsep}{3pt}
	\begin{minipage}{.5\linewidth}
		\centering
		\begin{tabular}{cclcc}
			\toprule
            $n_s$ & $n_\rho$ & ~~$\hat{c}^{\ff}(\mu_0/2)$ & $\chi^2/\mathrm{d.o.f.}$ \\
			\midrule
			2 & 2 & 1.1324(64) & 19.77 / 15 \\
            \textbf{2} & \textbf{3} & \textbf{1.1213(74)} & \textbf{11.30 / 14} \\
			3 & 2 & 1.1093(97) & 10.31 / 14 \\
			3 & 3 & 1.112(11) & 10.03 / 13 \\
			\bottomrule
		\end{tabular}
	\end{minipage}%
	\begin{minipage}{.5\linewidth}
	\centering
	\begin{tabular}{cclcc}
		\toprule
        $n_s$ & $n_\rho$ & ~~$\hat{c}^{\kk}(\mu_0/2)$ & $\chi^2/\mathrm{d.o.f.}$ \\
		\midrule
		2 & 2 & 1.1670(55) & 15.95 / 15 \\
		\textbf{2} & \textbf{3} & \textbf{1.1586(63)} & \textbf{\phantom{ }8.96 / 14} \\
		3 & 2 & 1.1498(84) & \phantom{ }8.96 / 14 \\
		3 & 3 & 1.1532(96) & \phantom{ }8.43 / 13 \\
		\bottomrule
	\end{tabular}
\end{minipage}
    \caption{Different fits of the anomalous dimension and the resulting values
             of the running factor $\hat{c}(\mu_0/2)$, cf.~Eq.~\eqref{rgi_operator}.
             We quote results for both the \ff-scheme (left) and the
             \kk-scheme (right), with our preferred fit result highlighted.
     }
    \label{tab:sf_regime_fits}
\end{table}
To determine the anomalous dimension of the tensor current in the high-energy
regime we make use of a global fit procedure which combines the continuum
extrapolation at individual values of the strong coupling with a direct fit to
the anomalous dimension constrained by the expectation from perturbation
theory. Both our starting expression, and the fitting strategy, follow a
similar reasoning as the one discussed at length in Ref.~\cite{Campos:2018ahf}
for the similar case of the running quark mass.
Our starting expression to model $\SigT^\mathrm{I}$ is
\begin{align}
	\SigT^\mathrm{I}(u,a/L) &= 
    \exp\left(\int_{\sqrt{u}}^{\sqrt{\sigma(u)}}\mathrm{d}x
    \frac{\gamT(x)}{\beta(x)}\right)+\left(\sum_{n=2}^{n_\rho}\rho_n u^n\right)\bigg(\frac{a}{L}\bigg)^2 \,.
\end{align}
For the anomalous dimension, we use the asymptotic
expansions~\eqref{eq:asympt-gamma} as our ansatz,
\begin{align}
	\gamT(x) &= -x^2\sum_{n=0}^{n_s}t_nx^{2n} \,.
\end{align}
We try fitting with different values for $n_s$ and $n_\rho$ and find consistent
results as detailed in~\Cref{tab:sf_regime_fits}. For our final result, we
quote the fits with $n_s=2$, $n_\rho=3$. For the scheme labelled \ff the
parameters are given by
\begin{align}
	t_0^\ff &=  \frac{8}{3(4\pi)^2} \equiv \gamT^{(0)}       \,, &
    t_1^\ff &=  0.00627445          \equiv \gamT^{(1),\ff}   \,, &
    t_2^\ff &= +0.00073(25)                                  \,,
\intertext{with $\chi^2/\mathrm{d.o.f.}=0.807$, while for the scheme labelled \kk we obtain}
	t_0^\kk &= \frac{8}{3(4\pi)^2} \equiv \gamT^{(0)}       \,, & 
    t_1^\kk &=  0.00579501         \equiv \gamT^{(1),\kk}   \,, &
    t_2^\kk &= -0.00022(21)                                 \,,
\end{align}
with $\chi^2/\mathrm{d.o.f.}=0.640$.

\begin{figure}
	\centering
	\includegraphics[width=0.9\linewidth]{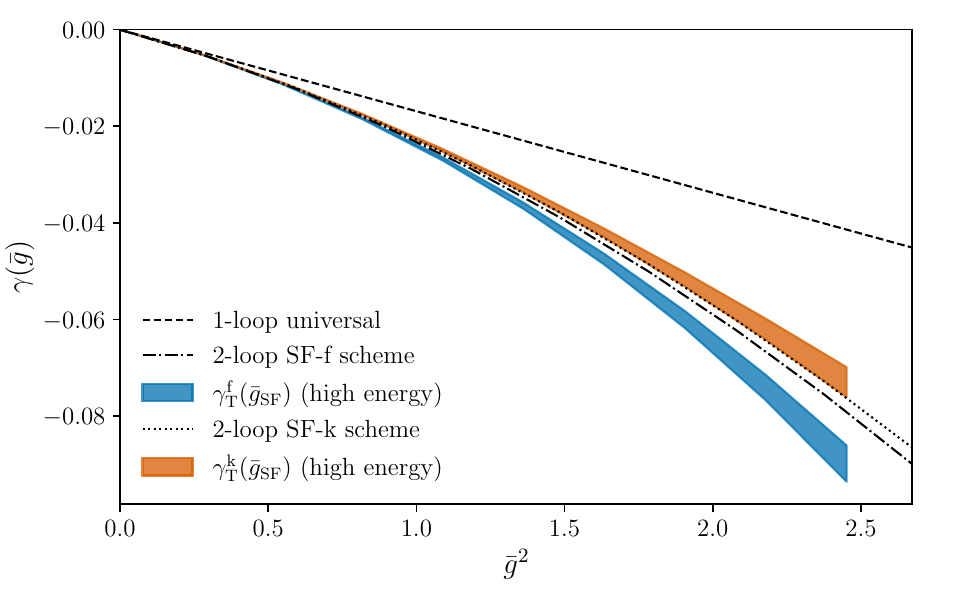}
    \caption{Non-perturbative anomalous dimension of the tensor current in the
             two SF schemes in the high-energy region. Perturbative
             expectations are shown for comparison.
            }
	\label{fig:tausfzt}
\end{figure}

In~\Cref{fig:tausfzt} the non-perturbative anomalous dimensions are compared to
the corresponding one-loop and two-loop perturbative predictions. For the
scheme labelled \kk the non-perturbative result agrees with the two-loop
prediction within errors while the discrepancy for the scheme labelled \ff
corresponds to several standard errors.

Having determined the anomalous dimension, which is constrained by construction
to make contact with two-loop perturbation theory at high energies, it is then
possible to determine directly the factor $\hat c(\mu_0/2) = \hat
c(\mu_{\rm\scriptscriptstyle PT})U(\mu_{\rm\scriptscriptstyle PT},\mu_0/2)$ in
Eq.~\eqref{eq:master}, in a way that makes it insensitive to any specific
prescription for $\mu_{\rm\scriptscriptstyle PT}$---see
Ref.~\cite{Campos:2018ahf} for details.
We quote for our two schemes
\begin{align}
\label{eq:high_running}
\hat c^{\mathrm{\ff}}(\mu_0/2) &= 1.1213(74) \,, &
\hat c^{\mathrm{\kk}}(\mu_0/2) &= 1.1586(63) \,.
\end{align}

\subsection{RG running at low energies}\label{sec:RG_uGF}

Below the matching scale $\mu_0/2$ we employ the GF scheme, for which we
have performed simulations at bare parameters such that the GF coupling is
close to one of the following seven values
\begin{align}
        \uu[GF] &\approx \{{2.12, 2.39, 2.73, 3.20, 3.86, 4.49, 5.29}\} \,.
\end{align}
These simulations are performed at three lattice spacings $L/a=8,12,16$, again
using non-perturbatively $\mathrm{O}(a)$ improved Wilson fermions but now with
a tree-level Symanzik-improved gauge action~\cite{Luscher:1984xn}. The value of
$\csw$ has been determined in Ref.~\cite{Bulava:2013cta}.  The chiral point is
tuned as in the SF regime via the PCAC relation in Eq.~\eqref{eq:pcac_mass},
with the corresponding non-perturbative value of $\cA$~\cite{Bulava:2015bxa}.
In this case, our non-perturbative results for $\cT$ from
Section~\ref{sec:improvement} are actually utilised to $\rmO(a)$ improve the
tensor current.  All computations are carried out at fixed topological charge
$Q=0$ by projecting onto the trivial topological sector, as explained in
Ref.~\cite{DallaBrida:2016kgh}.

\subsubsection{Boundary improvement}\label{ssec:boundary_improvement}

%
One relevant source of systematic uncertainty comes from the fact that the
$\mathrm{O}(a)$ improvement coefficients $\ct$ and $\cttil$, associated to
boundary counterterms in the SF setup, are only known within perturbation
theory.
For the tree-level Symanzik-improved gauge action they are actually only
known to one-loop from Refs.~\cite{Takeda:2003he,Vilaseca}.
While the impact of the perturbative truncation is negligible at small values
of $u$, it may become relevant in the low-energy region of the running.
This effect was studied in Ref.~\cite{Campos:2018ahf} for the case of the
running mass, and turned out to be negligible within statistical uncertainties
in the computation of the renormalisation constant $\ZP$.
However, in the case of the dependence on $\cttil$, to which fermionic
correlation functions are most sensitive, an accidental cancellation pushes the
perturbative truncation effect on $\ZP$ one order further in $g_0^2$.
This does not happen in the case of $\ZT$, meaning that we have to reassess the
issue here.

\begin{figure}[t]
	\centering
	\includegraphics[width=0.9\linewidth]{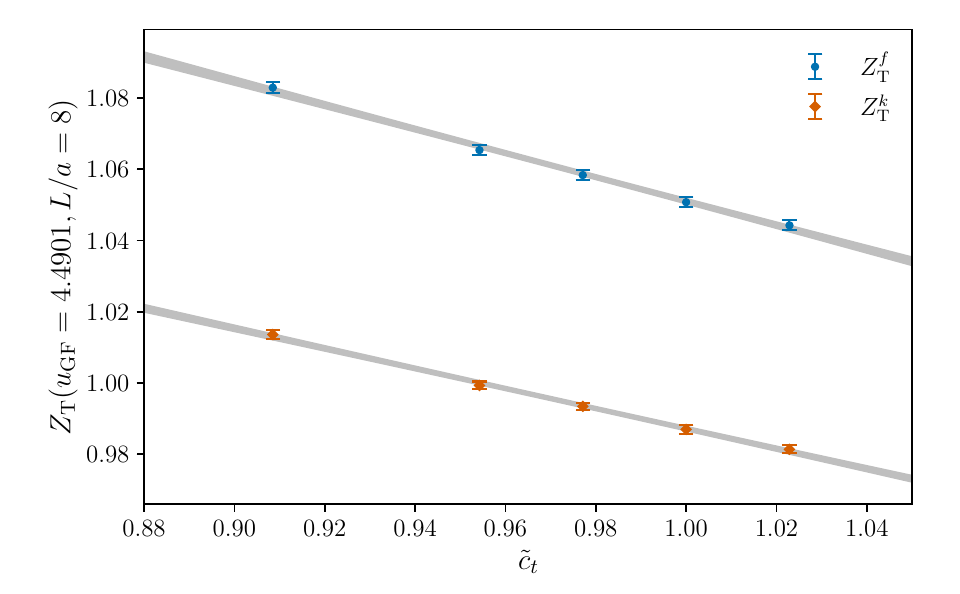}
    \caption{Dependence of the tensor renormalisation constant $\ZT$ on
             $\cttil$ for the \ff-scheme (circles) and the \kk-scheme
             (diamonds) at $\uGF=4.4901$ on $L/a=8$ lattices. 
            }
	\label{fig:cttilde_sys}
\end{figure}

To that effect, we have performed additional simulations at $u = 4.4901$ and
$L/a=8$, where we vary $\ct$ and $\cttil$ independently.
For $\ct$ we find a very mild dependence on the value used in the simulation,
and proceed to neglect that source of uncertainty.
However, for $\cttil$ we find a fairly strong dependence, as can be seen
from~\Cref{fig:cttilde_sys}.
In order to account for this effect by including an estimate of the systematic
uncertainty incurred in, we follow the same reasoning as in
Ref.~\cite{Campos:2018ahf}: linear error propagation suggests that the effect
of a shift $\delta\cttil$ on the value of $\ZT$ will have the form
\begin{gather}
	\label{eq:cttilsys}
	\delta_{\cttil}\ZT \approx \left|\frac{\partial \ZT}{\partial \cttil}\right|\delta \cttil \,.
\end{gather}
The slope $\partial \ZT/\partial \cttil$ at $u = 4.4901$ can be extracted from
a linear fit to our $L/a=8$ data, as depicted in~\Cref{fig:cttilde_sys}.
In order to estimate the effect at a different value of $u$ and/or $L/a$, we
posit the scaling law
\begin{gather}
	\frac{\partial \ZT}{\partial \cttil} \approx \xi u\frac{a}{L} \,,
\end{gather}
where $\xi$ is some constant coefficient.
The rationale is that the slope has a leading behaviour proportional to~$g_0^2$
in perturbation theory, and, the effect being associated to an $\mathrm{O}(a)$
improvement counterterm, it is expected to vanish linearly in $a$ at small
values of the lattice spacing.
By applying this to the result of our linear fit, we estimate:
\begin{gather}
\xi^\ff=-0.601(25)\,,\quad\xi^\kk=-0.502(19)\,.
\end{gather}
Finally, in order to apply Eq.~\eqref{eq:cttilsys} we conservatively use a
value of the shift $\delta \cttil$ corresponding to 100\% of the one-loop
perturbative deviation from the tree-level value $\cttil=1$.
Note that, at the level of the step-scaling function $\SigT$, our modelling of
the uncertainty leads to
\begin{align}
  \frac{\delta_{\cttil}\SigT}{\SigT} &\approx
     \left|\frac{\sigma(u)}{2\ZT(2L)}-\frac{u}{\ZT(L)}\right|
     |\xi|\frac{a}{L}\delta \cttil \,,
\end{align}
which implies that the uncertainty affecting the values of $\ZT$ that enter the ratio
undergoes a partial cancellation, making $\SigT$ less sensitive to this effect than
$\ZT$ itself.

The resulting systematic uncertainty induced in $\SigT$ is quoted as the second
number in parentheses in the relevant tables of Appendix~\ref{app:running}.
Note that it is largely subdominant with respect to the statistical
uncertainty, save for a few $L/a=8~\to~16$ steps where it is still smaller but
of comparable size.
In the case of $\ZT$, on the other hand, the systematic uncertainty can be
sizeable, which will be commented upon below when the matching at a hadronic
scale is discussed.

\subsubsection{Determination of the anomalous dimension}

\begin{figure}
	\centering
	\includegraphics[width=0.9\linewidth]{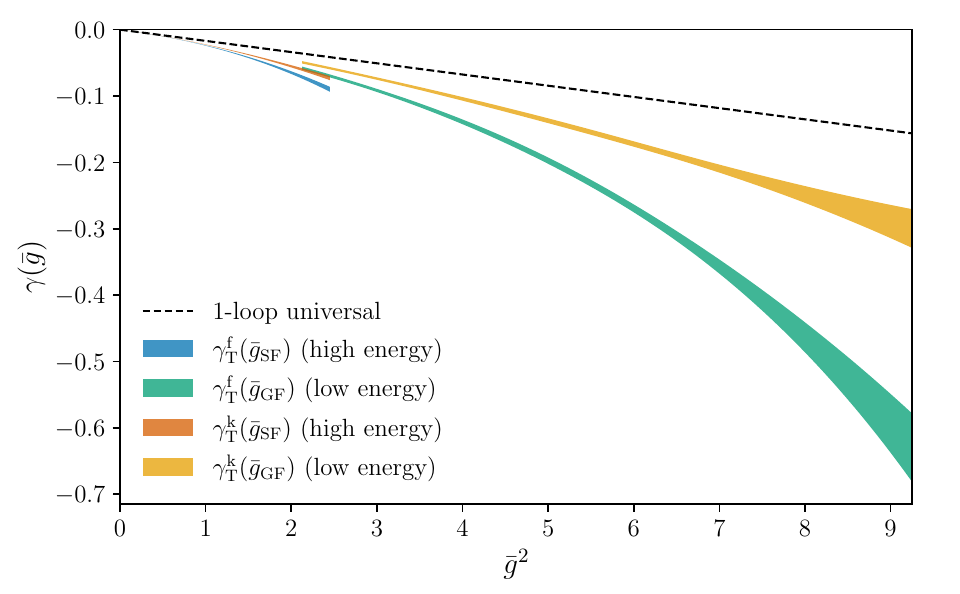}
    \caption{Non-perturbative anomalous dimension of the tensor current in the
             two schemes. The perturbative one-loop expectation is shown for
             comparison.  See~\Cref{fig:tausfzt} for a close up of the SF
             region.
            }
	\label{fig:taufullzt}
\end{figure}

To determine the anomalous dimension in the low-energy region we once again
make use of a global fit procedure, which in this case is strictly necessary as
the values of the GF coupling $\uu[GF]$ are not exactly tuned to a constant for
different values of $L/a$.  The ratios of RG functions are parametrised as
\begin{align}
	f(x) &= \frac{\gamT(x)}{\beta(x)} = \frac{1}{x}\sum_{n=0}^{n_r}f_nx^{2n} \,,
\end{align}
and we perform a global fit to the relation 
\begin{align}
	\SigT(u,a/L) &= \exp\left(\int_{\sqrt{u}}^{\sqrt{\Sigma(u,a/L)}} 
        \mathrm{d}x f(x)\right)+\left(\sum_{n=n_{\rho,\mathrm{start}}}^{n_{\rho,\mathrm{stop}}}\rho_nu^n\right)(a/L)^2 \\
	             &= \exp\left(\bigg[f_0\log(x)+\sum_{n=1}^{n_r}f_n
        \frac{x^{2n}}{2n}\bigg]_{\sqrt{u}}^{\sqrt{\Sigma(u,a/L)}}\right)+\left(\sum_{n=n_{\rho,\mathrm{start}}}^{n_{\rho,\mathrm{stop}}}\rho_nu^n\right)(a/L)^2 \,,
\end{align}
to obtain the parameters $f_n$.  For our best fits with $n_r=2$,
$n_{\rho,\mathrm{start}}=1$ and $n_{\rho,\mathrm{stop}}=2$ we obtain the
running factors
\begin{align}
\label{eq:low_running}
U^\ff(\mu_0/2,\muhad) &= 0.6475(59) \,, &
U^\kk(\mu_0/2,\muhad) &= 0.7519(45) \,.
\end{align}
With these parameters, we can also reconstruct the anomalous dimension via the relation
\begin{align}
	\gamT(\bar{g}) &= 
    -\bar{g}^2\frac{\sum_{n=0}^{n_r}f_n\bar{g}^{2n}}{\sum_{k=0}^{k_t}p_k\bar{g}^{2n}} \,,
\end{align}
where the $\beta$-function is parametrised as in Eq.~(4.12) of
Ref.~\cite{DallaBrida:2016kgh}.
For our best fits we obtain
\begin{align}
	f_0^\ff &= 0.326(63)    \,, &
    f_1^\ff &= 0.050(31)    \,, &
    f_2^\ff &=+0.0035(34)   \,,
\intertext{with $\chi^2/\mathrm{d.o.f.}=0.831$, and}
	f_0^\kk &= 0.298(47)    \,, &
    f_1^\kk &= 0.043(22)    \,, &
    f_2^\kk &=-0.0021(24)   \,,
\end{align}
with $\chi^2/\mathrm{d.o.f.}=1.130$. The corresponding covariance matrices can
be found in~\Cref{app:cov_low_energy}. The curves obtained from these fits are
shown in~\Cref{fig:taufullzt}, together with those derived in the high-energy
regime, and with the one-loop prediction.

The overall effect of the systematic error related to $\cttil$ to the total
squared error of the running factor in the low-energy regime, taken into
account via the procedure described in Section~\ref{ssec:boundary_improvement},
corresponds to about $4\%$ for both schemes.

\subsection{Matching at a hadronic scale}\label{sec:had_match}

As the final step in our renormalisation strategy, we need to compute the
renormalisation constant $\ZT$ at a fixed hadronic scale $\muhad=1/\Lhad$ for
changing bare couplings $g_0^2$ which match the couplings used in large-volume
simulations. In practice, we aim at the interval $\beta\in[3.40, 3.85]$ used by
the CLS consortium. The hadronic scale is fixed by a $L/a=20$ lattice with
$\beta=3.79$, resulting in $\uhad=9.25$. Using the scale setting of
Ref.~\cite{Bruno:2016plf}, this corresponds to an energy scale
$\muhad=233(8)\,\MeV$.  Lattices with $L/a=24,16,12,10$ were then tuned to
match this scale covering the range of CLS couplings.

\begin{table}[t!]
	\centering
	\begin{tabular}{lrlllll}
		\toprule
		{$L/a$} & {$\beta$} & {$\kappa$} & {$\uGF$} & {$Lm$} & {$\ZT^\ff(g_0^2, a\muhad)$} & {$N_{\rm ms}$} \\
		\midrule
		10 & 3.400000 & 0.136804 & 9.282(40) & $-0.0236(35)  $ & 1.308(6)(15)   & 2489 \\
		10 & 3.411000 & 0.136765 & 9.290(31) & $+0.0189(22)  $ & 1.276(4)(15)   & 4624 \\
		12 & 3.480000 & 0.137039 & 9.417(43) & $-0.0115(31)  $ & 1.328(7)(12)   & 1868 \\
		12 & 3.488000 & 0.137021 & 9.393(40) & $+0.0035(24)  $ & 1.329(6)(12)   & 2667 \\
		12 & 3.497000 & 0.137063 & 9.118(51) & $-0.0102(28)  $ & 1.318(8)(12)   & 1491 \\
		16 & 3.629800 & 0.137163 & 9.638(35) & $-0.0062(12)  $ & 1.3955(81)(90) & 6362 \\
		16 & 3.649000 & 0.137158 & 9.417(36) & $-0.0024(14)  $ & 1.4016(91)(88) & 4837 \\
		16 & 3.657600 & 0.137154 & 9.169(43) & $-0.0039(15)  $ & 1.3818(90)(85) & 3262 \\
		16 & 3.671000 & 0.137148 & 9.045(54) & $+0.0009(24)  $ & 1.371(12)(8)   & 1553 \\
		20 & 3.790000 & 0.137048 & 9.256(36) & $-0.00053(99) $ & 1.4105(75)(66) & 4305 \\
		24 & 3.893412 & 0.136894 & 9.370(61) & $-0.00001(100)$ & 1.471(14)(5)   & 3008 \\
		24 & 3.912248 & 0.136862 & 9.132(49) & $+0.00010(76) $ & 1.430(12)(5)   & 5086 \\
		\bottomrule
	\end{tabular}
	\caption{Results for $\ZT$ in the hadronic matching region (\ff-scheme).
	The first error is statistical, the second is systematic due to the use of a
	one-loop result for the improvement coefficient $\cttil$.}
	\label{tab:Zmatchf}
\end{table}
\begin{table}[t!]
	\centering
	\begin{tabular}{lrlllll}
		\toprule
		{$L/a$} & {$\beta$} & {$\kappa$} & {$\uGF$} & {$Lm$} & {$\ZT^\kk(g_0^2, a\muhad)$} & {$N_{\rm ms}$} \\
		\midrule
		10 & 3.400000 & 0.136804 & 9.282(40) & $-0.0236(35)  $ & 1.109(4)(12)   & 2489 \\
		10 & 3.411000 & 0.136765 & 9.290(31) & $+0.0189(22)  $ & 1.089(3)(12)   & 4624 \\
		12 & 3.480000 & 0.137039 & 9.417(43) & $-0.0115(31)  $ & 1.116(4)(10)   & 1868 \\
		12 & 3.488000 & 0.137021 & 9.393(40) & $+0.0035(24)  $ & 1.117(3)(10)   & 2667 \\
		12 & 3.497000 & 0.137063 & 9.118(51) & $-0.0102(28)  $ & 1.1141(44)(98) & 1491 \\
		16 & 3.629800 & 0.137163 & 9.638(35) & $-0.0062(12)  $ & 1.1546(32)(75) & 6362 \\
		16 & 3.649000 & 0.137158 & 9.417(36) & $-0.0024(14)  $ & 1.1586(37)(73) & 4837 \\
		16 & 3.657600 & 0.137154 & 9.169(43) & $-0.0039(15)  $ & 1.1527(39)(71) & 3262 \\
		16 & 3.671000 & 0.137148 & 9.045(54) & $+0.0009(24)  $ & 1.1452(48)(70) & 1553 \\
		20 & 3.790000 & 0.137048 & 9.256(36) & $-0.00053(99) $ & 1.1697(37)(55) & 4305 \\
		24 & 3.893412 & 0.136894 & 9.370(61) & $-0.00001(100)$ & 1.2060(66)(45) & 3008 \\
		24 & 3.912248 & 0.136862 & 9.132(49) & $+0.00010(76) $ & 1.1905(50)(44) & 5086 \\
		\bottomrule
	\end{tabular}
	\caption{Results for $\ZT$ in the hadronic matching region (\kk-scheme).
	The first error is statistical, the second is systematic due to the use of a
	one-loop result for the improvement coefficient $\cttil$.}
	\label{tab:Zmatchk}
\end{table}

\begin{figure}
	\centering
	\includegraphics[width=0.9\linewidth]{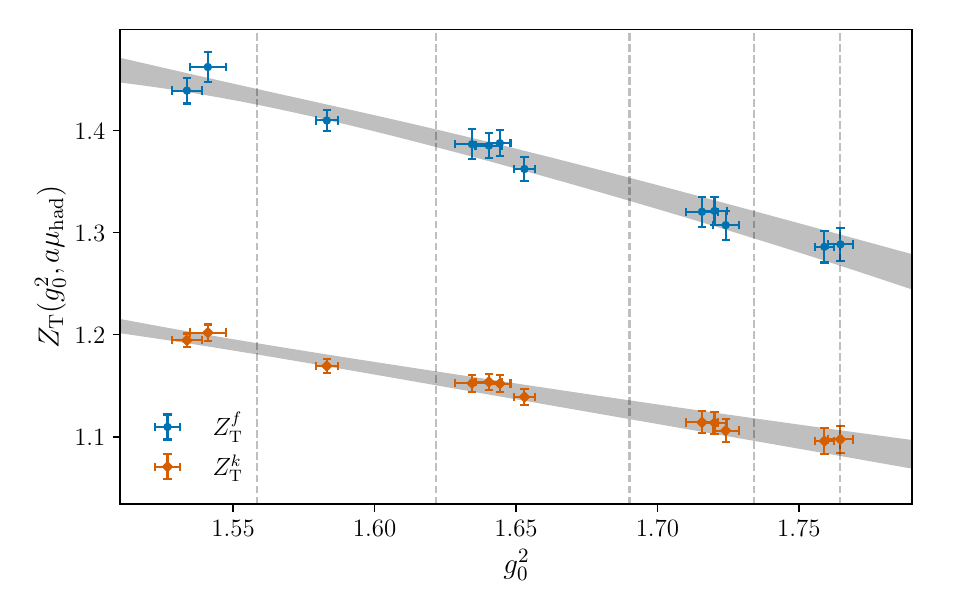}
    \caption{Tensor current renormalisation factor at the hadronic matching
             point $\uhad=9.25$ in the \ff-scheme (circles) and the \kk-scheme
             (diamonds). Dashed lines indicate the bare gauge couplings used in
             CLS simulations.
            }
	\label{fig:hadf}
\end{figure}

The full set of simulations in the hadronic regime is summarised in
Tables~\ref{tab:Zmatchf} and~\ref{tab:Zmatchk}. The tuning in both the coupling
$\uu[GF]$ and the mass $Lm$ is only precise up to a few standard deviations; we
account for this effect by performing a combined fit of the data as a function
of $g_0^2$, $\uu[GF]$ and $Lm$, in order to extract $\ZT(g_0^2,a\muhad)$ on our
line of constant physics defined by $\uhad=\bar{g}^2(\muhad)=9.25$, $Lm=0$. As
a model for our data we use the fit form
\begin{align}
\label{eq:ZThad}
\begin{split}
    \ZT(g_{0}^{2},\uu[GF],Lm) &= 
    \ZT(g_{0}^{2},a\muhad)+t_{10}\,(\uu[GF]-\uhad)+t_{01}\,Lm   \,, \\
    \ZT(g_{0}^{2},a\muhad)  &=
     z_{0}+z_{1}(\beta-\beta_{0})+z_{2}(\beta-\beta_{0})^{2}   \,.
\end{split}
\end{align}
For the free coefficients we obtain
\begin{align}
\label{eq:zmatchf}
	z_0^\ff &= 1.4178(75)   \,, &
    z_1^\ff &= 0.263(38)    \,, &
    z_2^\ff &=-0.21(10)    \,,
\end{align}
with $\chi^2/\mathrm{d.o.f.}=0.621$, and
\begin{align}
\label{eq:zmatchk}
	z_0^\kk &= 1.1748(58)   \,, &
    z_1^\kk &= 0.191(21)    \,, &
    z_2^\kk &=-0.043(55)    \,,
\end{align}
with $\chi^2/\mathrm{d.o.f.}=0.437$.
The corresponding functions and data points are presented in
Figure~\ref{fig:hadf} and their covariances can be found
in~\Cref{app:cov_hadronic_regime}.
In Table~\ref{tab:hadronic_Z_factor} we quote the values of  $\ZT$ at $\muhad$
for the values of $\beta$ where CLS ensembles have been simulated.

Note that, as hinted before, the systematic uncertainty due to the use of a
one-loop result for the improvement coefficient $\cttil$, that we
conservatively estimate via Eq.~\eqref{eq:cttilsys}, turns out to be
substantial.
It is indeed dominant with respect to the statistical uncertainty, except for
the largest $L/a=24$ lattices.
At the level of fit coefficients, the error on the zeroth fit parameter is
dominated by the systematic error estimate (77\% for $\ff$ and 88\% for $\kk$)
while the contribution is subleading for the remaining parameters (10\% and 5\%
for $\ff$, 24\% and 11\% for $\kk$).
Note also that systematic uncertainties are $100\%$ correlated by construction,
and should be treated in that way when the values of $\ZT$ quoted are employed.

\subsection{Total running and renormalisation factors}\label{sec:totalrun}

We are now in a position to quote our final results.
The total running factors relating RGI operator insertions to renormalised
insertions at $\muhad$ are given by the products of the two running factors in
Eqs.~\eqref{eq:high_running}~and~\eqref{eq:low_running}.
They are found to be
\begin{align}\label{eq:full_running}
	\hat c^\ff(\muhad) &= 0.7260(81)(14) \,, &
    \hat c^\kk(\muhad) &= 0.8711(70)(11) \,,
\end{align}
where the first error is statistical and the second is the systematic error
resulting from the fact that we only know the boundary $\mathrm{O}(a)$
improvement coefficients perturbatively.
We stress that these are continuum quantities, where the only dependence left
is in the renormalisation scheme.
We also stress that, as in the case of the values of $\ZT$ at $\muhad$,
systematic uncertainties are $100\%$ correlated by construction, and should be
treated in that way when the values of the above running factors are used.

\begin{figure}[t!]
	\centering
	\includegraphics[width=0.9\linewidth]{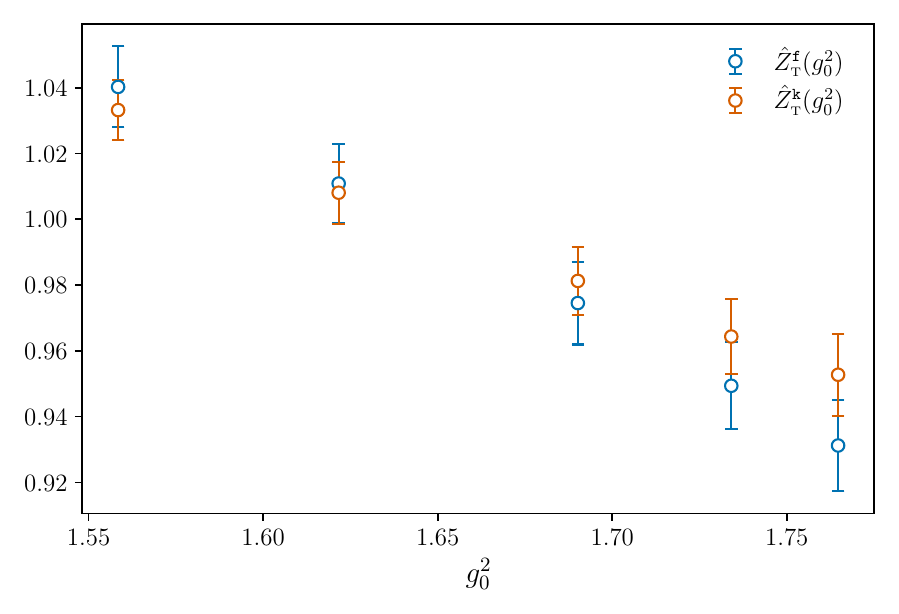}
    \caption{Comparison of the total renormalisation factors $\rZT(g_0^2)$ at
            the CLS couplings for both schemes. Note that the errors are highly
            correlated as all values for a given scheme share the running
            factor $\hat c$, whose uncertainty is dominant.
            }
	\label{fig:total_Z}
\end{figure}

By combining the running factors in Eq.~\eqref{eq:full_running} with the
renormalisation constants at $\muhad$ discussed in Section~\ref{sec:had_match},
it is furthermore possible to introduce a total renormalisation factor that
connects bare and RGI operator insertions, viz.
\begin{align}
    \rZT(g_0^2) &\equiv \hat c(\muhad) \ZT(g_{0}^{2},a\muhad)   \;.
\end{align}
This is a divergent quantity as $a \to 0$, which depends on the bare coupling
only, since the dependence on the hadronic matching scale $\muhad$ cancels by
construction up to cutoff effects.
By the same reason, and because the RGI is unique, the values of $\rZT$
computed through the two schemes can only differ by cutoff effects; in
particular, $\rZT^\ff(g_0^2)/\rZT^\kk(g_0^2) = 1 + \mathrm{O}(a^2)$ (up to
logarithmic corrections).
The value of $\rZT$ within the range in $g_0^2$ covered by our simulations can
be obtained trivially by multiplying the coefficients in
Eqs.~\eqref{eq:zmatchf} and~(\ref{eq:zmatchk}) by the corresponding factors in
Eq.~\eqref{eq:full_running}.
\Cref{fig:total_Z} shows a comparison of the total renormalisation factor in
both schemes as a function of the bare gauge coupling. 
The same comments as above regarding the correlation of systematic
uncertainties apply.

\begin{table}[t]
    \small
	\centering
	\renewcommand{\arraystretch}{1.25}
	\setlength{\tabcolsep}{3pt}
	\begin{tabular}{cll}
		\toprule
		$\beta$ & $\ZT^\ff(g_{0}^{2},a\muhad)$ 
                                 & $\ZT^\kk(g_{0}^{2},a\muhad)$ \\\midrule
         3.40   & 1.283(4)(15)   & 1.094(3)(12)     \\
         3.46   & 1.308(3)(13)   & 1.107(2)(11)     \\
         3.55   & 1.342(3)(11)   & 1.1265(18)(90)   \\
         3.70   & 1.3924(37)(78) & 1.1573(19)(65)   \\
         3.85   & 1.4328(50)(59) & 1.1861(25)(49)   \\
		\bottomrule
	\end{tabular}
    \caption{Results for $\ZT(g_{0}^{2},a\muhad)$ at CLS $\beta$-values for
             both schemes. The first error is statistical, the second is
             systematic due to the use of a one-loop result for the improvement
             coefficient $\cttil$.
            }
	\label{tab:hadronic_Z_factor}
\end{table}

\section{Conclusions}\label{sec:summary}

%
In the present work, we computed the renormalisation and running of non-singlet
quark bilinears with tensorial Lorentz structure, thereby addressing the last
missing non-trivial anomalous dimensions of two- and four-quark operators
within the ALPHA collaboration's $\Nf=3$ renormalisation programme.
Our approach, which is based on step scaling in the Schr\"odinger functional
and gradient flow schemes, allowed us to non-perturbatively compute the
operator anomalous dimension from the hadronic scale $\muhad=233(8)$ MeV all
the way up to electroweak energies, in two different renormalisation schemes.
As an accessory step, we computed non-perturbatively the improvement
coefficient $\cT$, required to obtain $\rmO(a)$ improved tensor currents in the
chiral limit, which is also relevant for computations of $\rmO(a)$ improved
amplitudes involving the latter.
In this procedure all error sources, statistical and systematic, are kept under
control.

The main results provided in the text are:
\begin{itemize}
\item
The non-perturbative values of the improvement coefficient $\cT(g_0^2)$ for a
non-perturbatively $\rmO(a)$ improved fermion action and a tree-level Symanzik
improved gauge action, in a large range of values of the bare gauge coupling
that includes those employed in large volume simulations (cf.
Eq.~\eqref{eq:interpolation_cT} and \Cref{tab:cT_cls}).
\item
The RG running factor connecting amplitudes of tensor currents at the hadronic
scale $\muhad$ and the corresponding RGI value, Eq.~\eqref{eq:full_running}.
These are continuum quantities, that only depend on the renormalisation scheme.
Results are provided in two different SF schemes for better control of the
systematics, with a ballpark 1\% precision. It is important to point out that
the running factors are in the continuum limit, and can therefore be applied to
continuum results obtained with any lattice action.
\item
The renormalisation constants of the improved currents in the range of
couplings relevant for CLS simulations, as well as the total renormalisation
factor relating bare and RGI hadronic matrix elements (cf.
Eq.~\eqref{eq:ZThad}, \Cref{tab:hadronic_Z_factor}). These results are
regularisation dependent, but still wide-ranging, since any computation based
on CLS ensembles can benefit from them.
\end{itemize}

Some interesting aspects of our results are worth stressing.
One is that the non-perturbative anomalous dimensions of tensor currents seem
to have generally larger values, and more sizeable deviations from low-order
perturbation theory, than the other independent anomalous dimension in the
two-quark sector---that is, the one of quark masses.
This makes an interesting potential case for the impact of non-perturbative
renormalisation on the systematic uncertainties of computations of tensor
amplitudes.
Another relevant observation is that a non-negligible source of uncertainty
comes from the lack of knowledge beyond one-loop perturbation theory about the
$\rmO(a)$ improvement coefficients related to the SF boundary.
This is a qualitative difference with respect to the computation of
renormalised quark masses in Ref.~\cite{Campos:2018ahf}, where an accidental
suppression first noticed in Ref.~\cite{Sint:1998iq} makes the relevant
renormalisation constants much less sensitive to the effect.
Efforts to suppress this source of uncertainty are thus part of the
methodological improvements required by future, higher-precision computations.

\begin{acknowledgement}%
J.H. wishes to thank the Yukawa Institute for Theoretical Physics, Kyoto
University, for its hospitality.  This work is supported by the Deutsche
Forschungsgemeinschaft (DFG) through the Research Training Group \textit{``GRK
2149: Strong and Weak Interactions -- from Hadrons to Dark Matter''} (J.H. and
F.J.). We acknowledge the computer resources provided by the WWU-IT of the
University of M\"unster (PALMA II) and thank its staff for support. F.J.~is
supported by UKRI Future Leader Fellowship MR/T019956/1. This project has
received funding from the European Union's Horizon 2020 research and innovation
programme under the Marie Skłodowska-Curie grant agreement No.  813942 (L.C.).
The work of C.P. and D.P. has been supported by the Spanish Research Agency
(Agencia Estatal de Investigación) through the grants IFT Centro de Excelencia
Severo Ochoa SEV-2016-0597 and CEX2020-001007-S and, grants FPA2015-68541-P,
PGC2018-094857-B-I00 and PID2021-127526NB-I00, all of which are funded by
MCIN/AEI/10.13039/501100011033. C.P. and D.P. also acknowledge support from the
project H2020-MSCAITN-2018-813942 (EuroPLEx) and the EU Horizon 2020 research
and innovation programme, STRONG-2020 project, under grant agreement No 824093.
The work of M.P. has been partially supported by the Italian PRIN ``Progetti di
Ricerca di Rilevante Interesse Nazionale -- Bando 2022'', prot. 2022TJFCYB and
by the Spoke 1 ``FutureHPC \& BigData'' of the Italian Research Centre in
High-Performance Computing, Big Data and Quantum Computing (ICSC), funded by
the European Union -- NextGenerationEU.
\end{acknowledgement}

\appendix
\section{Covariance matrices for fit parameters}\label{app:cov_matrices}

\subsection{Running at low energies}\label{app:cov_low_energy}

\begin{align}
	\cov(f_i^\ff, f_j^\ff) &={\begin{pmatrix}
			\begin{tabular}{@{\hspace{5pt}} S[table-format = +1.6e+1] @{\hspace{10pt}} S[table-format = +1.6e+1] @{\hspace{10pt}} S[table-format = +1.6e+1] @{\hspace{7pt}}}
			 3.985436e-03 & -1.884666e-03 &  2.017869e-04 \\
			-1.884666e-03 &  9.483937e-04 & -1.039767e-04 \\
			 2.017869e-04 & -1.039767e-04 &  1.187704e-05
			\end{tabular}
	\end{pmatrix}}\,,\\
	\cov(f_i^\kk, f_j^\kk) &={\begin{pmatrix}
			\begin{tabular}{@{\hspace{5pt}} S[table-format = +1.6e+1] @{\hspace{10pt}} S[table-format = +1.6e+1] @{\hspace{10pt}} S[table-format = +1.6e+1] @{\hspace{7pt}}}
			 2.182060e-03 & -1.004780e-03 &  1.040076e-04 \\
			-1.004780e-03 &  4.946745e-04 & -5.266863e-05 \\
			 1.040076e-04 & -5.266863e-05 &  5.713030e-06
			\end{tabular}
	\end{pmatrix}}\,.
\end{align}

\subsection{Matching at a hadronic scale}\label{app:cov_hadronic_regime}

\begin{align}
	\cov(z_i^\ff, z_j^\ff) &={\begin{pmatrix}
			\begin{tabular}{@{\hspace{5pt}} S[table-format = +1.6e+1] @{\hspace{10pt}} S[table-format = +1.6e+1] @{\hspace{10pt}} S[table-format = +1.6e+1] @{\hspace{7pt}}}
         5.671542e-05 & -4.288534e-05 & 1.410911e-04 \\
        -4.288534e-05 &  1.436836e-03 & 2.966136e-03 \\
         1.410911e-04 &  2.966136e-03 & 1.011492e-02
			\end{tabular}
	\end{pmatrix}}\,,\\
    \cov(z_i^\kk, z_j^\kk) &={\begin{pmatrix}
			\begin{tabular}{@{\hspace{5pt}} S[table-format = +1.6e+1] @{\hspace{10pt}} S[table-format = +1.6e+1] @{\hspace{10pt}} S[table-format = +1.6e+1] @{\hspace{7pt}}}
		 3.351458e-05 & -4.802689e-05 & 9.309619e-05 \\
		-4.802689e-05 &  4.458109e-04 & 6.658459e-04 \\
		 9.309619e-05 &  6.658459e-04 & 3.038084e-03
	\end{tabular}
\end{pmatrix}}\,.
	\end{align}

\setlength{\tabcolsep}{6pt} 
\section{Computation of $\cT$}\label{app:improvement}
%
This appendix collects the simulation parameters of the Schr\"odinger
functional gauge field ensembles employed for the determinations of the tensor
current improvement coefficient $\cT$ (Table~\ref{tab:gauge_parameters}) as
well as the associated results for the PCAC quark mass $am$ and $\cT$ from the
variants of Ward identity extractions discussed in
Section~\ref{sec:improvement} (Table~\ref{tab:cT_results}).
\begin{table}[h!]
    \small
	\centering
	\renewcommand{\arraystretch}{1.25}
	\setlength{\tabcolsep}{8pt}
	\begin{tabular}{llllrcl}
		\toprule
		ID       & $L^3\times T/a^4$   & $\beta$ & $\kappa$ &   MDU &$P(Q=0)$ & $\tau_\mathrm{exp}$[MDU]   \\
		\midrule
		A1k1     & $12^3\times 17$     & 3.3   &   0.13652  & 20480 & 0.365 & 8.33(46)   \\
		A1k3     & $12^3\times 17$     & 3.3   &   0.13648  &  6876 & 0.357 & 8.33(46)   \\
		A1k4     & $12^3\times 17$     & 3.3   &   0.1365   & 96640 & 0.366 & 8.33(46)   \\
		\midrule
		E1k1     & $14^3\times 21$     & 3.414 &   0.1369   & 38400 & 0.353 & 10.2(8)    \\
		E1k2     & $14^3\times 21$     & 3.414 &   0.13695  & 57600 & 0.375 & 10.2(8)    \\
		\midrule
		B1k1     & $16^3\times 23$     & 3.512 &   0.137    & 20480 & 0.389 & 22.2(3.3)  \\
		B1k2     & $16^3\times 23$     & 3.512 &   0.13703  &  8192 & 0.341 & 22.2(3.3)  \\
		B1k3     & $16^3\times 23$     & 3.512 &   0.1371   & 16384 & 0.458 & 22.2(3.3)  \\
		B1k4     & $16^3\times 23$     & 3.512 &   0.13714  & 27856 & 0.402 & 22.2(3.3)  \\
		\midrule
		C1k1     & $20^3\times 29$     & 3.676 &   0.1368   &  7848 & 0.334 & 63(17)     \\
		C1k2     & $20^3\times 29$     & 3.676 &   0.137    & 15232 & 0.450 & 63(17)     \\
		C1k3     & $20^3\times 29$     & 3.676 &   0.13719  & 15472 & 0.645 & 63(17)     \\
		\midrule
		D1k2     & $24^3\times 35$     & 3.81  &   0.13701  &  6424 & 0.457 & 154(31)    \\
		D1k4     & $24^3\times 35$     & 3.81  &   0.137033 & 85008 & 0.696 & 154(31)    \\
		\bottomrule
	\end{tabular}
        \caption{Simulation parameters' summary for the gauge field
            configuration ensembles labeled by `ID'.  MDU denotes the total
            length of the Markov chain for each ensemble in molecular dynamics
            units. $P(Q_0)$ gives the fraction of configurations, for which the
            topological charge $Q$ vanishes. $\tau_\mathrm{exp}$ is the
            exponential autocorrelation time used for the tail in the
            statistical data analysis (cf.~Ref.~\cite{Schaefer:2010hu}), which
            is estimated from the integrated autocorrelation time of the
            correlation function $f_1$ on the longest Markov chain for each
            value of $\beta$.  All measurements are separated by $8$ MDU except
            for ensembles A1k3 (4) and D1k4 (16).  The range of lattice
            spacings covered by the ensembles D through A is
            $0.042\,\fm\lesssim a\lesssim 0.105\,\fm$. 
            }
	\label{tab:gauge_parameters}
\end{table}

\begin{table}[h!]
    \small
	\centering
	\renewcommand{\arraystretch}{1.25}
	\setlength{\tabcolsep}{8pt}
	\begin{tabular}{llll>{\sffamily\bfseries}ll}
	\toprule
		ID   & $am$           & $\cT^{\Xstyle\lbrace T/4\rbrace}$   
                                            & $\cT^{\Xstyle \lbrace T/3 \rbrace}$   
                                                          & $\cTalt^{\Xstyle\lbrace T/4 \rbrace}$   
                                                                         & $\cTalt^{\Xstyle\lbrace T/3 \rbrace}$   \\
	\midrule
		A1k1 & $-0.00287(61)$ & 0.117(10)   & 0.114(8)    & 0.146(10)    & 0.143(8)    \\
		A1k3 & $+0.00105(95)$ & 0.132(13)   & 0.114(9)    & 0.160(13)    & 0.142(9)    \\
		A1k4 & $-0.00119(33)$ & 0.113(6)    & 0.114(4)    & 0.142(6)     & 0.143(4)    \\
		     & $0           $ & 0.121(8)    & 0.114(5)    & 0.150(8)     & 0.143(5)    \\
	\midrule
		E1k1 & $+0.00270(20)$ & 0.105(5)    & 0.088(4)    & 0.118(5)     & 0.101(4)    \\
		E1k2 & $+0.00042(13)$ & 0.108(5)    & 0.089(4)    & 0.123(5)     & 0.105(4)    \\
		     & $0           $ & 0.108(6)    & 0.089(5)    & 0.124(6)     & 0.105(5)    \\
	\midrule
		B1k1 & $+0.00552(20)$ & 0.084(5)    & 0.076(5)    & 0.085(5)     & 0.077(5)    \\
		B1k2 & $+0.00435(28)$ & 0.075(10)   & 0.072(8)    & 0.078(10)    & 0.075(8)    \\
		B1k3 & $+0.00157(18)$ & 0.099(7)    & 0.088(5)    & 0.107(7)     & 0.096(5)    \\
		B1k4 & $-0.00056(16)$ & 0.093(5)    & 0.074(4)    & 0.102(5)     & 0.083(4)    \\
		     & $0           $ & 0.094(4)    & 0.078(3)    & 0.103(4)     & 0.087(3)    \\
	\midrule
		C1k1 & $+0.01322(17)$ & 0.070(6)    & 0.059(5)    & 0.044(6)     & 0.032(5)    \\
		C1k2 & $+0.00601(11)$ & 0.072(5)    & 0.066(4)    & 0.061(5)     & 0.055(4)    \\
		C1k3 & $-0.00110(11)$ & 0.063(5)    & 0.061(4)    & 0.066(5)     & 0.063(4)    \\
		     & $0           $ & 0.065(4)    & 0.062(3)    & 0.066(4)     & 0.063(3)    \\
	\midrule
		D1k2 & $+0.00073(15)$ & 0.052(7)    & 0.052(4)    & 0.051(8)     & 0.051(6)    \\
		D1k4 & $-0.00007(3) $ & 0.055(2)    & 0.049(2)    & 0.056(3)     & 0.049(2)    \\
		     & $0           $ & 0.055(2)    & 0.049(2)    & 0.055(2)     & 0.049(2)    \\
	\bottomrule
	\end{tabular}
        \caption{Results for the PCAC quark mass $am$ and different
          determinations of the tensor current improvement coefficient $\cT$,
          as described in Section~\ref{sec:improvement}, on the individual
          gauge field ensembles of Table~\ref{tab:gauge_parameters} and in the
          chiral limit ($am\to 0$).  The errors of individual ensemble results
          are statistical, while the ones in the chiral limit follow from the
          orthogonal distance regression procedure of Ref.~\cite{Boggs1989}.
          Our preferred determination $\cTalt^{\Xstyle\lbrace T/4 \rbrace}$ is
          highlighted in boldface.
          }
	\label{tab:cT_results}
\end{table}

\clearpage

\section{Computation of $\ZT$}\label{app:running}

\begin{table}[h!]
	\centering
	\setlength{\tabcolsep}{6pt}
	\begin{tabular}{lrlllll}
\toprule
        \multicolumn{1}{c}{$u$} 
                 & \multicolumn{1}{c}{$L/a$} 
                      & \multicolumn{1}{c}{$\beta$} 
                                 & \multicolumn{1}{c}{$\kappa$} 
                                            & \multicolumn{1}{c}{$\ZT^{\ff}(L/a)$} 
                                                          & \multicolumn{1}{c}{$\ZT^{\ff}(2L/a)$} 
                                                                        & \multicolumn{1}{c}{$\SigT^{\ff}(L/a)$} \\
\midrule
		1.110000 & 6  & 8.540300 & 0.132336 & 0.98266(36) & 0.99491(44) & 1.02140(59) \\
		1.110000 & 8  & 8.732500 & 0.132134 & 0.98490(34) & 1.00060(77) & 1.02145(86) \\
		1.110000 & 12 & 8.995000 & 0.131862 & 0.99054(54) & 1.01015(94) & 1.0224(11)  \\
\midrule
		1.184400 & 6  & 8.217000 & 0.132690 & 0.98413(38) & 0.99453(46) & 1.02008(62) \\
		1.184400 & 8  & 8.404400 & 0.132477 & 0.98528(38) & 1.0034(10)  & 1.0242(11)  \\
		1.184400 & 12 & 8.676900 & 0.132172 & 0.99164(63) & 1.0131(10)  & 1.0244(12)  \\
\midrule
		1.265600 & 6  & 7.909100 & 0.133057 & 0.98417(40) & 0.99643(53) & 1.02265(68) \\
		1.265600 & 8  & 8.092900 & 0.132831 & 0.98595(40) & 1.00415(92) & 1.0248(10)  \\
		1.265600 & 12 & 8.373000 & 0.132492 & 0.99432(65) & 1.0152(11)  & 1.0240(13)  \\
\midrule
		1.362700 & 6  & 7.590900 & 0.133469 & 0.98440(42) & 0.99982(60) & 1.02668(76) \\
		1.362700 & 8  & 7.772300 & 0.133228 & 0.98703(43) & 1.0098(13)  & 1.0299(14)  \\
		1.362700 & 12 & 8.057800 & 0.132854 & 0.99313(71) & 1.0203(13)  & 1.0306(15)  \\
\midrule
		1.480800 & 6  & 7.261800 & 0.133934 & 0.98478(46) & 1.00221(68) & 1.02970(85) \\
		1.480800 & 8  & 7.442400 & 0.133675 & 0.98821(47) & 1.01361(80) & 1.03314(95) \\
		1.480800 & 12 & 7.729900 & 0.133264 & 0.99550(76) & 1.0255(12)  & 1.0336(14)  \\
\midrule
		1.617300 & 6  & 6.943300 & 0.134422 & 0.98740(50) & 1.00684(69) & 1.03284(88) \\
		1.617300 & 8  & 7.125400 & 0.134142 & 0.98899(49) & 1.0173(14)  & 1.0367(15)  \\
		1.617300 & 12 & 7.410700 & 0.133699 & 1.00020(94) & 1.0318(18)  & 1.0355(20)  \\
\midrule
		1.794300 & 6  & 6.605000 & 0.134983 & 0.98851(58) & 1.01104(97) & 1.0375(12)  \\
		1.794300 & 8  & 6.791500 & 0.134677 & 0.99287(55) & 1.0258(17)  & 1.0422(18)  \\
		1.794300 & 12 & 7.068800 & 0.134209 & 1.0034(10)  & 1.0423(19)  & 1.0430(22)  \\
\midrule
		2.012000 & 6  & 6.273500 & 0.135571 & 0.99359(64)  & 1.0206(11) & 1.0437(13)  \\
		2.012000 & 8  & 6.468000 & 0.135236 & 0.99594(65)  & 1.0374(13) & 1.0519(15)  \\
		2.012000 & 12 & 6.729950 & 0.134760 & 1.00663(100) & 1.0557(16) & 1.0536(19)  \\
		2.012000 & 16 & 6.934600 & 0.134412 & 1.02042(91)  & 1.0702(24) & 1.0515(25)  \\
\bottomrule
	\end{tabular}
    \caption{Results for step scaling of $\ZT^{\ff}$ in the high-energy region.
            }
\end{table}

\begin{table}[h!]
	\centering
	\setlength{\tabcolsep}{6pt}
	\begin{tabular}{lrlllll}
\toprule
        \multicolumn{1}{c}{$u$} 
                 & \multicolumn{1}{c}{$L/a$} 
                      & \multicolumn{1}{c}{$\beta$} 
                                 & \multicolumn{1}{c}{$\kappa$} 
                                            & \multicolumn{1}{c}{$\ZT^{\kk}(L/a)$} 
                                                          & \multicolumn{1}{c}{$\ZT^{\kk}(2L/a)$} 
                                                                        & \multicolumn{1}{c}{$\SigT^{\kk}(L/a)$} \\
\midrule
		1.110000 & 6  & 8.540300 & 0.132336 & 0.96712(31) & 0.97995(38) & 1.02104(51) \\
		1.110000 & 8  & 8.732500 & 0.132134 & 0.97094(30) & 0.98628(64) & 1.02048(73) \\
		1.110000 & 12 & 8.995000 & 0.131862 & 0.97784(47) & 0.99617(80) & 1.02089(96) \\
\midrule
		1.184400 & 6  & 8.217000 & 0.132690 & 0.96694(33) & 0.97824(39) & 1.01997(53) \\
		1.184400 & 8  & 8.404400 & 0.132477 & 0.96993(32) & 0.98748(88) & 1.02310(97) \\
		1.184400 & 12 & 8.676900 & 0.132172 & 0.97778(53) & 0.99729(87) & 1.0222(11)  \\
\midrule
		1.265600 & 6  & 7.909100 & 0.133057 & 0.96528(34) & 0.97841(44) & 1.02249(59) \\
		1.265600 & 8  & 8.092900 & 0.132831 & 0.96913(34) & 0.98685(78) & 1.02363(88) \\
		1.265600 & 12 & 8.373000 & 0.132492 & 0.97856(56) & 0.99798(98) & 1.0223(12)  \\
\midrule
		1.362700 & 6  & 7.590900 & 0.133469 & 0.96347(36) & 0.97949(50) & 1.02623(65) \\
		1.362700 & 8  & 7.772300 & 0.133228 & 0.96829(36) & 0.9896(11)  & 1.0278(12)  \\
		1.362700 & 12 & 8.057800 & 0.132854 & 0.97642(59) & 1.0008(11)  & 1.0276(13)  \\
\midrule
		1.480800 & 6  & 7.261800 & 0.133934 & 0.96152(39) & 0.97923(57) & 1.02887(73) \\
		1.480800 & 8  & 7.442400 & 0.133675 & 0.96726(39) & 0.99067(66) & 1.03050(80) \\
		1.480800 & 12 & 7.729900 & 0.133264 & 0.97665(61) & 1.00320(97) & 1.0301(12)  \\
\midrule
		1.617300 & 6  & 6.943300 & 0.134422 & 0.96084(42) & 0.98022(57) & 1.03162(75) \\
		1.617300 & 8  & 7.125400 & 0.134142 & 0.96540(40) & 0.9910(11)  & 1.0334(13)  \\
		1.617300 & 12 & 7.410700 & 0.133699 & 0.97828(77) & 1.0063(15)  & 1.0318(17)  \\
\midrule
		1.794300 & 6  & 6.605000 & 0.134983 & 0.95796(48) & 0.98000(79) & 1.03577(98) \\
		1.794300 & 8  & 6.791500 & 0.134677 & 0.96539(46) & 0.9949(13)  & 1.0382(15)  \\
		1.794300 & 12 & 7.068800 & 0.134209 & 0.97859(86) & 1.0123(16)  & 1.0380(18)  \\
\midrule
		2.012000 & 6  & 6.273500 & 0.135571 & 0.95747(52) & 0.98284(88) & 1.0409(11)  \\
		2.012000 & 8  & 6.468000 & 0.135236 & 0.96411(53) & 0.9998(10)  & 1.0457(12)  \\
		2.012000 & 12 & 6.729950 & 0.134760 & 0.97782(81) & 1.0186(12)  & 1.0457(15)  \\
		2.012000 & 16 & 6.934600 & 0.134412 & 0.99194(73) & 1.0329(19)  & 1.0435(21)  \\
\bottomrule
	\end{tabular}
    \caption{Results for step scaling of $\ZT^{\kk}$ in the high-energy region.
            }
\end{table}

\begin{table}[h!]
	\centering
	\setlength{\tabcolsep}{4pt}
	\begin{tabular}{llrlllll}
		\toprule
		\multicolumn{1}{c}{$u$} 
                   & \multicolumn{1}{c}{$\Sigma_u(L/a)$} 
                                & \multicolumn{1}{c}{$L/a$} 
                                     & \multicolumn{1}{c}{$\beta$} 
                                                & \multicolumn{1}{c}{$\kappa$}
                                                           & \multicolumn{1}{c}{$\ZT^{\ff}(L/a)$}
                                                                         & \multicolumn{1}{c}{$\ZT^{\ff}(2L/a)$} 
                                                                                      & \multicolumn{1}{c}{$\SigT^{\ff}(L/a)$} \\
\midrule
		2.1293(24) & 2.4226(49) & 8  & 5.371500 & 0.133621 & 1.00494(58) & 1.0365(19) & 1.0314(20)(12) \\
		2.1213(21) & 2.5049(70) & 12 & 5.543070 & 0.133314 & 1.01804(72) & 1.0603(29) & 1.0415(29)(8) \\
		2.1257(25) & 2.5356(57) & 16 & 5.700000 & 0.133048 & 1.02923(87) & 1.0706(27) & 1.0402(28)(5) \\
\midrule
		2.3910(26) & 2.7722(63) & 8  & 5.071000 & 0.134217 & 1.00956(69) & 1.0490(22) & 1.0391(23)(15) \\
		2.3919(25) & 2.8985(82) & 12 & 5.242465 & 0.133876 & 1.02424(89) & 1.0765(33) & 1.0510(33)(9) \\
		2.3900(30) & 2.9375(70) & 16 & 5.400000 & 0.133579 & 1.0373(11)  & 1.0886(23) & 1.0494(25)(6) \\
\midrule
		2.7353(31) & 3.2650(79) & 8  & 4.764900 & 0.134886 & 1.01494(73) & 1.0640(25) & 1.0484(26)(17) \\
		2.7371(38) & 3.406(11)  & 12 & 4.938726 & 0.134508 & 1.0341(14)  & 1.0953(34) & 1.0592(36)(11) \\
		2.7359(35) & 3.485(11)  & 16 & 5.100000 & 0.134169 & 1.0484(12)  & 1.1160(33) & 1.0645(34)(7) \\
\midrule
		3.2046(37) & 3.968(11)  & 8  & 4.457600 & 0.135607 & 1.02479(96) & 1.0909(36) & 1.0645(36)(21) \\
		3.2051(47) & 4.174(13)  & 12 & 4.634654 & 0.135200 & 1.0437(15)  & 1.1266(48) & 1.0795(48)(13) \\
		3.2029(52) & 4.263(15)  & 16 & 4.800000 & 0.134821 & 1.0649(16)  & 1.1482(41) & 1.0782(42)(9) \\
\midrule
		3.8619(45) & 5.070(16)  & 8  & 4.151900 & 0.136326 & 1.0381(12)  & 1.1315(39) & 1.0900(39)(26) \\
		3.8725(60) & 5.389(23)  & 12 & 4.331660 & 0.135927 & 1.0648(21)  & 1.1815(64) & 1.1096(64)(16) \\
		3.8643(63) & 5.485(21)  & 16 & 4.500000 & 0.135526 & 1.0861(18)  & 1.2284(62) & 1.1310(60)(11) \\
\midrule
		4.4870(56) & 6.207(23)  & 8  & 3.947900 & 0.136747 & 1.0584(13)  & 1.1791(51) & 1.1141(49)(31) \\
		4.4945(76) & 6.785(34)  & 12 & 4.128217 & 0.136403 & 1.0858(23)  & 1.2638(74) & 1.1640(72)(19) \\
		4.4901(77) & 6.890(47)  & 16 & 4.300000 & 0.136008 & 1.1159(22)  & 1.291(11)  & 1.1567(99)(12) \\
\midrule
		5.3040(88) & 7.953(44)  & 8  & 3.754890 & 0.137019 & 1.0825(19)  & 1.280(12)  & 1.183(11)(4) \\
		5.300(11)  & 8.782(47)  & 12 & 3.936816 & 0.136798 & 1.1221(27)  & 1.428(25)  & 1.273(23)(2) \\
		5.301(13)  & 9.029(61)  & 16 & 4.100000 & 0.136473 & 1.1511(36)  & 1.433(16)  & 1.244(15)(1) \\
		\bottomrule
	\end{tabular}
    \caption{Results for step scaling of $\ZT^{\ff}$ in the low-energy region.
            The notation $\Sigma_u$ refers to the value of the coupling in the
            $2L/a$ lattice.
            }
\end{table}

\begin{table}[h!]
	\centering
	\setlength{\tabcolsep}{4pt}
	\begin{tabular}{llrlllll}
		\toprule
		\multicolumn{1}{c}{$u$} 
                   & \multicolumn{1}{c}{$\Sigma_u(L/a)$} 
                                & \multicolumn{1}{c}{$L/a$} 
                                     & \multicolumn{1}{c}{$\beta$} 
                                                & \multicolumn{1}{c}{$\kappa$} 
                                                           & \multicolumn{1}{c}{$\ZT^{\kk}(L/a)$} 
                                                                         & \multicolumn{1}{c}{$\ZT^{\kk}(2L/a)$} 
                                                                                      & \multicolumn{1}{c}{$\SigT^{\kk}(L/a)$} \\
\midrule
		2.1293(24) & 2.4226(49) & 8  & 5.371500 & 0.133621 & 0.97888(47) & 1.0055(15) & 1.0272(16)(11) \\
		2.1213(21) & 2.5049(70) & 12 & 5.543070 & 0.133314 & 0.99238(58) & 1.0270(22) & 1.0349(23)(6)  \\
		2.1257(25) & 2.5356(57) & 16 & 5.700000 & 0.133048 & 1.00324(70) & 1.0387(21) & 1.0354(22)(5)  \\
\midrule
		2.3910(26) & 2.7722(63) & 8  & 5.071000 & 0.134217 & 0.97952(56) & 1.0120(18) & 1.0332(19)(12) \\
		2.3919(25) & 2.8985(82) & 12 & 5.242465 & 0.133876 & 0.99398(72) & 1.0361(25) & 1.0424(26)(8)  \\
		2.3900(30) & 2.9375(70) & 16 & 5.400000 & 0.133579 & 1.00709(88) & 1.0472(20) & 1.0398(21)(5)  \\
\midrule
		2.7353(31) & 3.2650(79) & 8  & 4.764900 & 0.134886 & 0.97949(61) & 1.0183(19) & 1.0396(20)(15) \\
		2.7371(38) & 3.406(11)  & 12 & 4.938726 & 0.134508 & 0.9985(11)  & 1.0464(27) & 1.0480(29)(9)  \\
		2.7359(35) & 3.485(11)  & 16 & 5.100000 & 0.134169 & 1.01212(93) & 1.0642(25) & 1.0514(27)(6)  \\
\midrule
		3.2046(37) & 3.968(11)  & 8  & 4.457600 & 0.135607 & 0.98135(78) & 1.0322(27) & 1.0519(28)(18) \\
		3.2051(47) & 4.174(13)  & 12 & 4.634654 & 0.135200 & 1.0010(12)  & 1.0617(31) & 1.0606(34)(11) \\
		3.2029(52) & 4.263(15)  & 16 & 4.800000 & 0.134821 & 1.0203(12)  & 1.0809(29) & 1.0594(31)(7)  \\
\midrule
		3.8619(45) & 5.070(16)  & 8  & 4.151900 & 0.136326 & 0.98448(94) & 1.0506(27) & 1.0672(28)(22) \\
		3.8725(60) & 5.389(23)  & 12 & 4.331660 & 0.135927 & 1.0096(15)  & 1.0869(46) & 1.0766(48)(13) \\
		3.8643(63) & 5.485(21)  & 16 & 4.500000 & 0.135526 & 1.0296(13)  & 1.1274(43) & 1.0950(44)(9)  \\
\midrule
		4.4870(56) & 6.207(23)  & 8  & 3.947900 & 0.136747 & 0.9934(11)  & 1.0700(32) & 1.0770(33)(25) \\
		4.4945(76) & 6.785(34)  & 12 & 4.128217 & 0.136403 & 1.0185(17)  & 1.1292(50) & 1.1088(52)(14) \\
		4.4901(77) & 6.890(47)  & 16 & 4.300000 & 0.136008 & 1.0455(16)  & 1.1508(62) & 1.1007(62)(9)  \\
\midrule
		5.3040(88) & 7.953(44)  & 8  & 3.754890 & 0.137019 & 1.0028(14)  & 1.1163(51) & 1.1133(53)(29) \\
		5.300(11) & 8.782(47)   & 12 & 3.936816 & 0.136798 & 1.0368(20)  & 1.1907(91) & 1.1484(90)(16) \\
		5.301(13) & 9.029(61)   & 16 & 4.100000 & 0.136473 & 1.0611(24)  & 1.2034(68) & 1.1341(68)(10) \\
\bottomrule
	\end{tabular}
    \caption{Results for step scaling of $\ZT^{\kk}$ in the low-energy region.
            The notation $\Sigma_u$ refers to the value of the coupling in the
            $2L/a$ lattice.
            }
\end{table}

\small
\addcontentsline{toc}{section}{References}
\bibliographystyle{JHEP}
\bibliography{mainbib}

\end{document}